\documentclass[preprint,11pt]{elsarticle}
\usepackage[utf8]{inputenc}
\usepackage{amsfonts, amssymb, amsmath, amsthm,mathtools}
\usepackage{mathrsfs}
\usepackage{graphicx,graphics}
\usepackage{placeins}
\usepackage{epstopdf}
\usepackage{multicol}
\usepackage{multirow}
\usepackage{subfig}
\usepackage{algorithm}
\usepackage{algorithmic}
\usepackage{setspace}
\usepackage{anysize}
\usepackage{color}
\usepackage{calrsfs}
\usepackage{hyperref}
\marginsize{2cm}{2cm}{2cm}{2cm} 
\begin{document}

\title{Methodology to estimate ionospheric scintillation risk maps and
their contribution to position dilution of precision on the ground}

\author[ADD1]{Alexandra Koulouri}
\author[] {Nathan D. Smith}
\author[ADD3]{Bruno C. Vani}
\author[ADD4]{Ville Rimpil\"aine}
\author[ADD2]{Ivan Astin}
\author[ADD2]{Biagio Forte}
\address[ADD1]{Laboratory of Mathematics, Tampere University of Technology, P.O.\ Box 692, 33101 Tampere, Finland}
\address[ADD2]{ Department of Electronic and Electrical Engineering, University of Bath, BA2 7AY, Bath, UK}
\address[ADD4]{Department of Physics, University of Bath, BA2 7AY,  Bath, UK}
\address[ADD3]{Universidade Estadual Paulista J{\'u}lio de Mesquita
Filho}



\begin{abstract}
Satellite-based communications, navigation systems and many
scientific instruments rely on observations of trans-ionospheric
signals. The quality of these signals can be deteriorated by
ionospheric scintillation which can have detrimental effects on the
mentioned applications.  Therefore, monitoring of ionospheric
scintillation and quantifying its effect on the ground are of
significant interest. In this work, we develop a methodology which
estimates the scintillation induced ionospheric uncertainties in the
sky and translates their impact to the end-users on the ground.
First, by using the risk concept from decision theory and by
exploiting the intensity and duration of scintillation events (as
measured by the $S_4$ index), we estimate ionospheric risk maps that
could readily give an initial impression on the effects of
scintillation on the satellite-receiver communication. However, to
better understand the influence of scintillation on the positioning
accuracy on the ground, we formulate a new weighted dilution of
precision (WPDOP) measure that incorporates the ionospheric
scintillation risks as weighting factors for the given
satellite-receiver constellations. These weights depend implicitly
on scintillation intensity and duration thresholds which can be
specified by the end-user based on the sensitivity of the
application, for example.  We demonstrate our methodology by using
scintillation data from South America, and produce ionospheric risk
maps which illustrate broad scintillation activity, especially at
the equatorial anomaly. Moreover, we construct ground maps of WPDOP
over a grid of hypothetical receivers which reveal that ionospheric
scintillation can also affect such regions of the continent that are
not exactly under the observed ionospheric scintillation structures.
Particularly, this is evident in cases when only the Global
Positioning System (GPS) is available.

\end{abstract}


\maketitle

\section{Introduction}
\label{intro}

Many industrial and scientific applications exploit Global
Navigation Satellite Systems (GNSS), that includes Global
Positioning System (GPS), Globalnaya Navigazionnaya Sputnikovaya
Sistema, (GLONASS) and Galileo (European navigation system), for
precision positioning, agriculture, transportation, surveying,
construction and geodesy among others. Satellite links using
frequencies up to few GHz (VHF through L-band) can experience
significant signal amplitude and phase fluctuations due to
 small-scale plasma density inhomogeneities, a phenomenon known as
 scintillation \citep{Crane1977,Aarons1982}. {The amplitude scintillation (i.e. temporal fluctuation of the
signal strength) and phase scintillation occur often at lower
latitudes and can cause periods of reduced signal power at the
receiver's antenna (fading)}. Therefore, GNSS signal-to-noise-ratio
may drop below the receiver-tracking threshold \citep{Jiao2015}
thereby leading to
 loss of positioning
information, that can persist for a significant period of time
\citep{Aquino2005}.

Many characteristics of scintillation are well-studied
\citep{Aarons1982,Basu1988,Conker2003,Li2010,Beniguel2011,Jiao2015,Forte2017,Sahithi2019}.
It has been shown that scintillation activity varies with operating
frequency, local time, season, geomagnetic activity level(planetary
K-index, $K_p$), and the 11-year solar cycle
\citep{Skone2001b,Li2008,Olatunbosun2017,Basu1988,Li2010,Jiao2015}.
Moreover, the scintillation effects vary according to the user's
location (i.e. geomagnetic latitude). During high scintillation
activity, these effects can be a serious problem for GNSS users in
certain regions, such as polar and equatorial latitudes, while users
in other regions (e.g. middle latitudes) are not affected to a large
extent \citep{Spogli2009,Abadi2014,Jiao2015,Luo2018}.
 Equatorial regions experience strong
scintillation mostly after sunset until a few hours after midnight
local time \citep{Paul2010,Jiao2015,Jiao2016}. 
In equatorial ionosphere, the most significant scintillation
activity occurs with deep signal fades causing a GNSS receiver to
lose lock and degrade positioning accuracy
\citep{SKONE2001,Aquino2005,Moreno2011,Marques2015}.

Limited case studies have been carried out related to the effects of
scintillation for the end-user \citep{Zumberge1997,
Moreno2011,Sreeja2011,Xu2012,Pi2017,Alfonsi2018,Luo2018,O.Moraes2018,Guo2019}
and often global climatological models fail to describe extreme
day-to-day variability of scintillation \citep{Wernik2003}.
Moreover, because scintillation is frequency dependent and cannot be
eliminated, for example by combining observable signals at different
frequencies \citep{Mainul2012,IJssel2016}, the constant monitoring
of scintillation activity using in-situ measurements alongside with
the development of mathematical methodologies, which describe,
quantify and interpret its impact on GNSS communications, is
substantial to mitigate its effect. In this framework, we develop a
mathematical methodology to estimate ionospheric scintillation risks
based on scintillation measurements and decision theory. These risks
are given by the joint probability of scintillation intensity and
duration estimated using information obtained from ground based
scintillation monitors. Thus, in contrast to the commonly estimated
statistics (e.g. averages or percentages of scintillation
measurements above a threshold as in
\citep{Spogli2009,Li2010,Sreeja2011,Jacobsen2014}), we use data
collected from a network of scintillation monitors to construct
ionospheric risk maps that describe and monitor
scintillation behavior. While the ionospheric maps can readily be
used to get an (initial) impression of the effects of scintillation
on a satellite-receiver communication, in this work, we propose to
incorporate the scintillation risk into the receiver's position
dilution of precision (PDOP) \citep{Langley1999}.

Standard PDOP is a well-known way to quantify the reliability and
integrity of a GNSS positioning system by using the geometry of the
available satellite constellations
\citep{Sairo2003,Zhang2013,Chen2013,Tahsin2015,Teng2018}. Here,
 we show how the degradation of PDOP
occurs relative to the location of ionization structures in the
F-region due to scintillation. To do that, we propose to use a
weighted position dilution of precision (WPDOP) which is estimated
by assigning different weights to receiver-satellite links using the
ionospheric scintillation risks along the corresponding
line-of-sights. Hence, in addition to the number and the geometry of
the visible satellites, we take into account the effect of
scintillation. Previous works have employed empirical error tracking
models, e.g. Conker model \citep{Conker2003}, to implicitly
introduce underlying scintillation activity by weighting receiver's
measurements with the inverse of the variance of the tracking error.
However, these approaches are often susceptible to model limitations
and requirements of heuristically selected parameters and their
applicability is limited due to the requirement of specific
software/hardware that is available mainly in operating ground
stations and not in standard GNSS receivers
\citep{Aquino2009,Silva2010}. Instead, our approach utilizes
weighing factors based on risks associated directly with monitored
scintillation data. With the help of these weights, we are able to
construct maps that describe the expected uncertainty in positioning
results during ionospheric scintillation. To the best of our
knowledge, this is the first time such a methodology is introduced.

The rest of the paper is organized as follows. In section~
\ref{sec:Theory}, we present the theoretical background and
mathematical expression of the risks and the proposed WPDOP. In
section~\ref{sec:material}, we describe the formulae and technical
details for the implementation of the proposed ionospheric risks and
the corresponding ground WPDOP values. In
section~\ref{section:results}, we demonstrate our methodology by
estimating ionospheric risk maps over South America using real
scintillation data. Subsequently, we construct ground maps based on
two different set-ups using either only GPS or GPS, GLONASS and
Galileo. Then, we discuss the main observations and further
potentials of the proposed methodology. Finally, in
section~\ref{sec:conclusions}, we present a summary of our work and
future aims.

\section{Methodology}\label{sec:Theory}
\subsection{Risk concept}\label{sec:preliminary}
While it may be straightforward to extract statistics (e.g. averages
or standard deviations) to describe scintillation \citep{Liu2015},
nevertheless it is important to interpret those statistics for an
end-user, in terms of the costs expected for a particular
application when scintillation is experienced. For this purpose, we
use risk analysis from decision theory \citep{Murphy2012}. Risk can
be defined as the expected loss incurred when an activity is
performed (for example satellite-receiver communication). To
estimate the risk we need i) to find features that describe the
condition in which we perform this activity (e.g. communication
under scintillation can be described through the measured $S_4$
value), ii) the frequency of occurrence of these features and iii)
how injurious the condition described by these features is to the
successful outcome of the activity. These are respectively expressed
by i) a feature vector ii) a probability and iii) a loss function.

In this section, we describe the background theory and definitions
about loss function and risk before making it application-specific
in the next section. Here, we denote random variables by capital
letters and their realizations by lowercase letters. We start with
the
 measurable feature $Z\in \mathcal{Z}$, where
$\mathcal{Z}$ represents the set of all possible feature vectors
(e.g. scintillation index, etc).

Let us introduce a 0-1 loss function
\begin{equation}\label{eq:binaryLoss}
\ell(z;z_\mathrm{th}) =
\begin{cases}
0 & \textrm{if $z<z_\mathrm{th}$}\\
1 &\textrm{otherwise},
\end{cases}
\end{equation}
which implies there is no loss when values of $z$ are less than a
specified threshold, but otherwise losses are equally injurious.
Although this is a simple loss function, it nevertheless is
instructive for this particular problem without increasing
computational complexity, and as will be shown it permits a
straightforward interpretation.

The risk is then defined as the expected value of the loss function
given by
\begin{equation}\label{eq:Risk_temp}
{r}(z_\mathrm{th}) =
\mathbb{E}[\ell(z;z_\mathrm{th})] = \int_{\mathcal{Z}}
\ell(z;z_\mathrm{th})\;p(z)\;\mathsf{d}z,
\end{equation}
where $\mathbb{E}[.]$ denotes the expectation value and $p(z)$ is
the probability density funciton of the measurable feature $Z$.
From (eq.~\ref{eq:binaryLoss}) and (eq.~\ref{eq:Risk_temp}), we have
that
\begin{equation}\label{eq:Risk_temp1}
{r}(z_\mathrm{th}) = \pi(Z\geq z_\mathrm{th}),
\end{equation}
where $\pi(.)$ denotes probability.  Thus, the risk has a simple
interpretation in terms of probability.

\subsection{Risk associated with ionospheric scintillation}
In this section, we apply the previous analysis to estimate risk
maps using scintillation features. Let $X$ denote a  scintillation
index (e.g. $S_4$ value). 
Moreover, we define as a \emph{scintillation event} a
time series of scintillation index values above a scintillation
threshold denoted by $x_\mathrm{th}$ over a duration limit
$d_\mathrm{th}$ (see Fig.~\ref{Fig:timeSeries}). The duration of
\emph{scintillation events} can be treated as a random variable
denoted by $D$. Based on the analysis of
section~\ref{sec:preliminary}, we have that $(X,D) \equiv Z$.

Using the insight that the vulnerability of a link depends on the
value of the scintillation index (e.g. strong scintillation when
$S_4>0.6$) and duration of a \emph{scintillation event}, we define
the loss function
\begin{equation}\label{eq:binaryLoss_new}
\ell(x,d;x_\mathrm{th},d_\mathrm{th}) =
\begin{cases}
0 &  \textrm{if $x< x_\mathrm{th}$ and $d< d_\mathrm{th}$}\\
1 &\textrm{otherwise},
\end{cases}
\end{equation}
where $x_\mathrm{th}$ and $d_\mathrm{th}$ are the intensity and
duration thresholds respectively. These thresholds can be chosen to
reflect the specifications of the receiver and the needs of safety
critical applications.

From (eq.~\ref{eq:Risk_temp1}), the risk due to scintillation
features is
\begin{equation}\label{eq:Risk}
r_\mathrm{ion}(x_\mathrm{th},d_\mathrm{th})=\pi(X\geq x_\mathrm{th},
D\geq d_\mathrm{th}).
\end{equation}
Hence, the risk of the underlying activity, depends on the joint
probability of the duration and intensity of \emph{scintillation
events} and takes values between 0 and 1.

\subsection{Effect of scintillation on the positioning accuracy}\label{sec:WPDOPtheory}
The proposed risk 
can readily be used to give an initial impression of the effect of
scintillation in a satellite-receiver communication. However, to
better understand the influence of scintillation on the positioning
accuracy on the ground, a well-known quantity as the dilution of
precision can be estimated \citep{Langley1999}.

Towards this aim, this section introduces a weighted position
dilution of precision (WPDOP) as a measure of uncertainty in
estimating the receiver's position. 
To put our analysis in perspective, we first discuss the theoretical
basis of the standard position dilution of precision (PDOP) from the
Bayesian point of view. Then, we explain how in addition to the
availability and geometry of the satellites, the proposed WPDOP can
include information about the effect of ionospheric scintillation.
\subsubsection{Standard position dilution of
precision}\label{sec:standardPDOP} Following similar analysis as in
\citep{Langley1999}, for a ground receiver at location
$\mathrm{r}=(r_x,r_y,r_z)^\mathrm{T}$, the positioning error at time
$t$, denoted by $\Delta \mathrm{r}=(\Delta r_x,\Delta r_y,\Delta
r_z)^\mathrm{T}$, can be expressed through the linear system
\begin{equation}\label{eq:linearSystem}
b= A\Delta\mathrm{r}+\varepsilon,
\end{equation}
where $b\in\mathbb{R}^S$ is a vector with the differences between
the measured and modelled pseudorange values,
 $A=[\mathrm{n}_1,\ldots,\mathrm{n}_S]^\mathrm{T}\in\mathbb{R}^{S\times 3}$ where $\mathrm{n}_s$ is a
 unit column
 vector pointing from the modelled (approximated) receiver position to the satellite
 and $S$ is the total number of visible
satellites at a time instance $t$.
 Error $\varepsilon\in\mathbb{R}^S$
represents the measurement noise plus model errors and ionospheric
effects (including scintillation). Further details about the
derivation of the aforementioned linear system are given in
Appendix~\ref{Appendix:PositioningError} \footnote{ We note here
that the clock offset is omitted (as a variable to be estimated)
from the current formulation. Matrix $A$, $\Delta\mathrm{r}$, $b$
and $\varepsilon$ are time-varying variables but for simplicity in
the notation, time $t$ has not been used in this context. Also,
linear system (eq.~\ref{eq:linearSystem}) helps us to introduce the
different dilution of precision formulae. The estimation of vector
$b$ is out of the scope of the current work, the interested reader
is referred to for example \citep{Teunissena}.}.

{In the PDOP standard probabilistic analysis
\citep{Langley1999,Misra2010}, the noise $\varepsilon$ is modelled
as i.i.d. Gaussian given by $\varepsilon\sim\mathcal{N}(0,\gamma
I^{S\times S})$ (with constant error variance $\gamma>0$ and
$I^{S\times S}$ identity matrix)}. A point estimate for the
positioning error $\Delta r$ can be obtained by solving
\begin{equation} \widehat{\Delta r}:=\mathrm{arg}\max_{\Delta
r}p(\Delta r|b), \end{equation}
 where
$p(\Delta r|b)$ is the posterior distribution of $\Delta r$ given by
$p(\Delta r|b)\propto p(b|\Delta r)\;\pi(\Delta r)$ based on the
Bayes theorem, where $p(b|\Delta r)\propto \exp{\{-\frac{1}{2\gamma}
(A\Delta r-b)^{\mathrm{T}}(A\Delta r-b)\}}$ is the likelihood and
$p(\Delta r)$ is the prior
distribution \citep{Kaipio2004}. 

With an uninformative prior $p(\Delta r)$, the positioning error can
be obtained by $\widehat{\Delta r}:=\mathrm{arg}\max_{\Delta
r}p(b|\Delta r)$
which is the maximum likelihood (ML) estimate. The ML estimate is  
\begin{equation}\label{eq:LS}
\widehat{\Delta \mathrm{r}}=(A^{\mathrm{T}}A)^{-1}A^{\mathrm{T}}b,
\end{equation}
and it is equal to the least squares solution of system (eq.~
\ref{eq:linearSystem}).
 The solution (eq.~\ref{eq:LS}) exists as long as there are 3 or more satellites
located in distinct directions in the sky. Quantitatively, the rank
of $(A^\mathrm{T}A)$ can be used as an indicator whether the problem
is well posed or not. For example, if
$\mathrm{rank}(A^\mathrm{T}A)\geq3$, then we can deduce that there
are enough satellites for the position estimation. However, the
accuracy of the solution  (eq.~\ref{eq:LS}) depends on the inverse
matrix $(A^{\mathrm{T}}A)^{-1}$ (whether it is well or
ill-conditioned). From the Bayesian point of view, this inverse
matrix multiplied by error variance $\gamma$ is the posterior
covariance of $p(\Delta r|b)$ given by
$\Gamma=\gamma(A^\mathrm{T}A)^{-1}$ and its diagonal elements
(posterior variances) can help to quantify the expected accuracy of
the estimate $\widehat{\Delta r}$ \citep{Kaipio2004}. Large
posterior variances may indicate that there is possibly an
overlapping  between satellites (eclipse phenomenon), collinearity
 or inappropriate geometric constellation.
The square root of the trace of the posterior $\Gamma$ normalized by
the constant variance $\gamma$  is the well-known position dilution
of precision (PDOP)\citep{Langley1999}. In particular, for a ground
receiver at location $\mathrm{r}$ and time $t$, the PDOP is given by
\begin{equation}\label{eq:PDOP}
\mathrm{PDOP}^{\mathrm{r},t} =
\sqrt{\frac{\mathrm{tr}(\Gamma)}{\gamma}}=
\sqrt{\mathrm{tr}\left((A^{\mathrm{T}}A)^{-1}\right)},
\end{equation}
where $\mathrm{tr}()$ denotes the trace of matrix given by
$\mathrm{tr}\left((A^{\mathrm{T}}A)^{-1}\right)=\sum_{s=1}^S\mathrm{diag}\left((A^{\mathrm{T}}A)^{-1}\right)$.
Hence, PDOP can be used to quantify the expected uncertainty in the
estimate (eq.~\ref{eq:LS}).

\subsubsection{Weighted position dilution of
precision}\label{sec:WPDOP_description}
Now, by considering the scintillation effect independently from all
the other errors, the noise term in the observation model
(eq.~\ref{eq:linearSystem}) can be decomposed into two uncorrelated
errors
\begin{equation}\label{eq:errordecomposition}
\varepsilon
=\varepsilon_\mathrm{sc}+\varepsilon_\mathrm{rem}\in\mathbb{R}^S.\end{equation}
{The error due to ionospheric scintillation can be approximated by a
Gaussian distribution given by}
$\varepsilon_\mathrm{sc}\sim\mathcal{N}(0,\Gamma_{\varepsilon_\mathrm{sc}})$
\citep{Misra2010} where its covariance is a diagonal matrix
$\Gamma_{\varepsilon_\mathrm{sc}}=\mathrm{diag}({\gamma_\mathrm{sc}^{(s)}})$
with the variances depending on the scintillation risks along the
available satellite-receiver ray-paths . Therefore, the error
variances due to scintillation can be expressed as a function of the
ionospheric risk $r^{(s)}_\mathrm{ion}$ along each ray-path $s$,
i.e.
\begin{equation}\label{eq:function_h}
\gamma_\mathrm{sc}^{(s)}= h(r^{(s)}_\mathrm{ion})\quad \mbox{for
}s=1,\ldots,S.
\end{equation}
Function $h(r_\mathrm{ion}^{(s)})$, since is the variance of the
measurement errors caused by scintillation in link $s$,  should be a
positive monotonically non decreasing function, and satisfy
$h(r_\mathrm{ion}^{(s)})\rightarrow \infty$ for risk
$r_\mathrm{ion}^{(s)}\rightarrow 1$ (more uncertainty is introduced
in the estimates due to scintillation) and
$h(r_\mathrm{ion}^{(s)})\rightarrow 0$  when risk $r_\mathrm{ion}^{(s)}\rightarrow 0$ (minimal effects due to scintillation in the estimates when risk is low).

{The remaining error $\varepsilon_\mathrm{rem}$, that encloses all
the other modelling and measurement related uncertainties, is
modelled as Gaussian, i.e. $\varepsilon_\mathrm{rem}\sim
\mathcal{N}(0,\gamma I^{S\times S})$ with $\gamma$ constant as in
the standard PDOP analysis}\footnote{{Gaussian distribution with
constant variance is a standard modelling assumption when prior
statistical information about the
observation error is limited or unavailable \citep{Misra2010}.}} \citep{Langley1999}.  
 We notice that in the current analysis the errors between different links are considered
uncorrelated which is a plausible assumption since there are usually
large distances between different ray-paths \footnote{{For GPS L1
frequency, the most effective spatial size of the electron density
irregularities that causes amplitude scintillation is  of order 400
m (first Fresnel scale at $\sim 350$ km altitude)
\citep{Rufenach1972,Kintner2005,Peng2019}.
The IPPs of two different ray-paths are separated by distances that
are much larger than the size of a single scintillation structure.
Therefore, it is very unlikely for two ray-paths to cross the same
ionospheric structure.}}.


Then, the covariance matrix of total error $\varepsilon$ given by
$\Gamma_\varepsilon=\gamma I^{S\times
S}+\Gamma_{\varepsilon_\mathrm{sc}}$ is a diagonal matrix
$\Gamma_\varepsilon=\mathrm{diag}(\gamma_{\varepsilon^{(s)}})$ with
elements
\begin{equation}
\label{eq:errorvariances}\gamma_{\varepsilon^{(s)}} = \gamma_{}+
h(r^{(s)}_\mathrm{ion})\quad \mbox{for }s=1,\ldots,S.
\end{equation}
The positioning error estimate  
is now
\begin{equation}
\label{eq:generalExpectedPositioningError}
\widehat{\Delta\mathrm{r}}=
(A^{\mathrm{T}}\Gamma_\varepsilon^{-1}A)^{-1}A^{\mathrm
T}\Gamma_\varepsilon^{-1} b, \end{equation}
 when $S\geq 3$ and
$\mathrm{rank}(A^{\mathrm{T}}\Gamma_\varepsilon^{-1}A)\geq 3$. The
posterior covariance of $\Delta \mathrm{r}$ is
\begin{equation}\label{eq:PostCovarianceWeights}
\Gamma = (A^\mathrm{T}\Gamma_{\varepsilon}^{-1}A)^{-1}
\end{equation}
and the weighted position dilution of precision (WPDOP) 
for a ground receiver at location $\mathrm{r}$ and time $t$ is
defined (in a similar way as in eq.~\ref{eq:PDOP}) as the square
root of the trace of the posterior $\Gamma$ normalized by the
constant $\gamma_\mathrm{}$ and it is given by
\begin{equation}\label{eq:WPDOP}
\mathrm{WPDOP}^{\mathrm{r},t} =
\sqrt{\frac{\mathrm{tr}(\Gamma_{})}{\gamma}}=\sqrt{\mathrm {tr}
\left(A^\mathrm{T}\left(\frac{\Gamma_\varepsilon}{\gamma}\right)^{-1}A\right)^{-1}}=
\sqrt{(\mathrm{tr}(A^\mathrm{T}WA)^{-1})},
\end{equation}
where
$\mathrm{tr}\left((A^{\mathrm{T}}WA)^{-1}\right)=\sum_{s=1}^S\mathrm{diag}\left((A^{\mathrm{T}}WA)^{-1}\right)$
and
$W=\left(\frac{\Gamma_\varepsilon}{\gamma_\mathrm{}}\right)^{-1}$ is
a diagonal matrix, with elements
\begin{equation}\label{eq:weights}
w_s=\frac{\gamma}{\gamma_{\varepsilon^{(s)}}}=\frac{\gamma}{\gamma+
h(r_\mathrm{ion}^{(s)})} \quad\textrm{ for}\quad s=1:S,
\end{equation}
where $0\leq w_s\leq1$ and $\gamma$ constant. {A rigorous derivation
of (eq.~\ref{eq:WPDOP}) is given in
appendix~\ref{sec:insightInDilutionOfPrecision}.}

Based on equation (eq.~\ref{eq:weights}) we have that $w_s=1$ when
$r_\mathrm{ion}^{(s)}=0$ (no scintillation, since
$h(r_\mathrm{ion}^{(s)})= 0$ ), and $w_s=0$ when
$r_\mathrm{ion}^{(s)}\rightarrow 1$ (if
$h(r_\mathrm{ion}^{(s)})\rightarrow \infty$). 
{The shape of $h()$ in (eq.~\ref{eq:weights}) may be selected based
on prior knowledge or application requirements.} In general, the
proposed WPDOP uses both information about the number of available
satellites, their geometry in the sky and the risks associated with
scintillation along the visible ray-paths. If
$\mathrm{rank}(A^\mathrm{T}WA)<3$ i.e. the number of available
satellites is less than 3 then $\mathrm{WPDOP}^{\mathrm{r},t}$ is
undefined.

\section{Methodology}\label{sec:material}

\subsection{Single shell ionospheric model}
The F-region (especially around 350 km of altitude) is of main
concern for GNSS users due to propagation errors caused by the
dynamics of the large electron density. To construct surface
ionospheric maps it is standard to consider the ionospheric thin
shell approximation at 350 km altitude, based on the assumption that
the entirety of the electron content (which affects a link) is
acting at the point where a ray-path of a satellite-receiver
intersects (or pierces) the ionospheric shell at that altitude
\citep{Davies1990}. Hence, an ionospheric risk map can be estimated
using ground measurements that are directly projected on the
ionospheric shell at the corresponding ionospheric pierce points
(IPPs) \footnote{IPP is referred to as the point where a
line-of-sight from the ground receiver to a satellite ``punctures''
the ionosphere at the ionospheric shell.} (similarly as in
\citep{Vani2017}). The projected (scintillation) measurements at the
350km ionospheric shell will be referred to as \emph{IPP data} in
the following text.
\subsection{Ionospheric computational domain and data
accumulation}\label{sec:CompDomainDataAcc} For the construction of
an ionospheric risk map, the thin ionospheric shell was discretized
using a uniform grid where each pixel got its own risk value. In the
results section, we estimated scintillation risk maps over South
America and as a reliable index of scintillation at low-latitudes,
we used $S_4$ data which is the ratio of the standard deviation of
the received signal intensity (measured by a 50 Hz sampling rate) to
the averaged intensity during an 1 minute
time interval. 
A risk was estimated at each ionospheric pixel $v$ based on the
joint probability (eq.~\ref{eq:Risk}) using \emph{IPP data}
accumulated over a time interval $\tau$. We note that there is an
implicit assumption of statistical stationarity i.e. the joint
probability does not change over time.

\subsection{A \emph{scintillation event}}
\begin{figure}[ht]
      \centering
          \includegraphics[width=0.5\columnwidth]{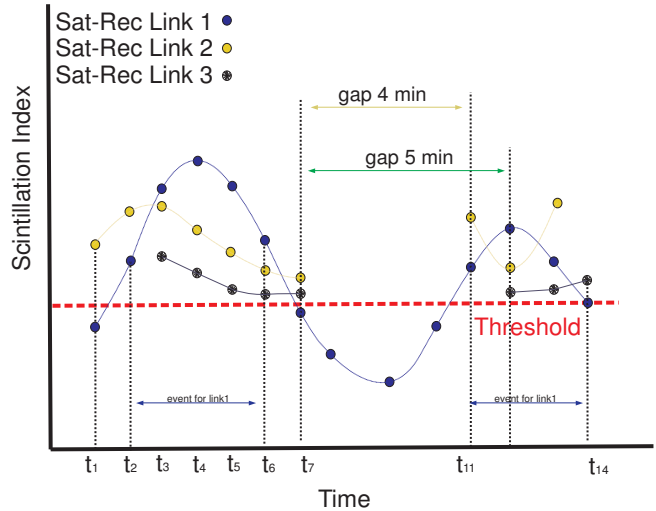}
      \caption{A schematic example which show how a \emph{scintillation event} is determined. This figure shows \emph{IPP data} samples
      (depicted with circles) in one ionospheric pixel over a short interval of time (the sampling period is considered 1 minute).
       Each colour corresponds to a data sequence from a different satellite-ground station link. 
       IPP values greater than the intensity threshold (depicted by the horizontal dashed
       line) indicate that the sample belongs to a \emph{scintillation
       event}.
       For link 1, we can observe that there are two distinct \emph{scintillation events} between $[t_2-t_6]$  and $[t_{11}-t_{14}]$.
       Links 2 and 3 have  data gaps of duration 4 and 5 minutes,
              respectively. If the gap threshold is 4 minutes then we can consider
               a single \emph{scintillation event} for link 2
      and  two scintillation events of short durations for Link 3.}
      \label{Fig:timeSeries}
\end{figure}

\label{subsect:scintillation
event} A \emph{scintillation event} was introduced briefly  in
section~\ref{sec:Theory}. Here, we explain more analytically the
properties of such events. Particularly, duration of a
\emph{scintillation event} is the time interval where the
scintillation values for a specific (satellite-scintillation
monitor) link are greater than a given threshold. In the schematic
example of Figure ~\ref{Fig:timeSeries}, we show how different
\emph{scintillation events} are determined using sequences of
\emph{IPP data} samples (depicted by small circles) for 3 different
links imagined that their ray-paths are passing through the same
ionospheric pixel $v$. The horizontal dashed line indicates a
selected intensity threshold $x_\mathrm{th}$. All the links
experience \emph{scintillation events} since there are data points
above the threshold. For link 1 (in blue), we can observe that there
are two distinctive scintillation events between $[t_2-t_6]$ and
$[t_{11}-t_{14}]$. However, for links 2 (in yellow) and 3 (in black)
we have to deal with data gaps related either to strong
scintillation or other reasons. In these cases, we have to decide
whether the available data and the corresponding gaps are considered
as a single scintillation event or not. In the current
implementation, we consider  that there is a single scintillation
event if the gap is small (i.e. less than 4 minutes). For a longer
gap, we consider that two separate scintillation events of shorter
durations take place.
Extra conditions based on the values of the scintillation samples
before and after the gap may also be applied to differentiate single
or multiple events. For example very high values of the $S_4$
samples on either side of a large gap (e.g. $S_4\geq 0.6$) can imply
a single event rather than two distinct events.

\subsection{Risk
estimation}\label{sec:EstimationOfProbabilities}
In this section we describe the mathematical formulae for the
estimation of the ionospheric risk maps. As we mentioned earlier,
risk maps are constructed by estimating pixel-wise the joint
probability (eq.~\ref{eq:Risk}) using the accumulated \emph{IPP
data} at each pixel of the discretized
 ionospheric domain over a period of time $\tau$. Based on that and (eq.~\ref{eq:Risk}) we can mathematically express  a
 risk in a pixel
as the conditional probability given pixel $v$ and time window
$\tau$, i.e.
\begin{equation}\label{eq:RiskInAPixel}
r^{v,\tau}_{\mathrm{ion}}(x_\mathrm{th},d_\mathrm{th}) = \pi(X\geq
x_\mathrm{th},D\geq d_\mathrm{th}|v,\tau).
\end{equation}
Moreover, we introduce the following notation for clarity. A
scintillation measurement is given by $x^{s}(\mathrm{v}(t))$, where
$\mathrm{v}(t)=(v_x(t),v_y(t))$ is the geographic location (at the
ionospheric pierce point) at time $t$. Super-script $s$ is used to
identify that the measured index corresponds to a particular
satellite-monitor link. Then, a \emph{scintillation event} is a time
series of scintillation samples $x^{s}(\mathrm{v}(t))\geq
x_{\mathrm{th}} $ for an interval $[t, t+\Delta d]$, where $d\geq
d_\mathrm{th}$, when these samples belong to a single pixel, $v$
(i.e. $\mathrm{v}(t)\in v$). Here, $\Delta$ is the sampling period
and $d_\mathrm{th}$ is an integer denoting the minimum number of
samples required to qualify for the event (a.k.a. duration
threshold). Thus, each \emph{IPP sample} can be represented pairwise
as $(x^{s}(\mathrm{v}(t)),d^{s}(t))$ that includes the scintillation
value and the number of samples in the scintillation event to which
$x^{s}(\mathrm{v}(t))$ belongs. {For example,  $d^{s}(t)=0$ when
$x^{s}(\mathrm{v}(t))<x_\mathrm{th}$ which means that sample
$x^{s}(\mathrm{v}(t))$ does not belong to a scintillation event. On
the other hand when $x^{s}(\mathrm{v}(t))\geq x_\mathrm{th}$ then
sample $x^{s}(\mathrm{v}(t))$ belongs to a scintillation event and
then $d^{s}(t)$ equals the number of consecutive samples (before and
after(and including) $x^{s}(\mathrm{v}(t))$ at time $t$) that exceed
threshold $x_\mathrm{th}$.}

In the following analysis, for simplicity in the probability
expressions, we write measurement $
x^{s}(\mathrm{v}(t))|_{t=t_j}=x^{s}[j] $, where $x^{s}[j]$ is the
value of the scintillation index at time $t_j$ for
satellite-receiver link $s$ and the corresponding IPP (ionospheric
location) of $x^{s}[j]$ is denoted by $\mathrm{v}^{s}[j]$.

The joint probability of $X$ and $D$  (eq.~\ref{eq:RiskInAPixel})
can be re-written
 as
\begin{equation}
\pi(X\geq x_{\mathrm{th}}, D\geq d_\mathrm{th}| v,\tau) = \pi( D\geq
d_\mathrm{th}| X\geq x_{\mathrm{th}}, v,\tau) \pi(X\geq
x_{\mathrm{th}}|v,\tau).
\end{equation}
The previous equation can be further expanded, i.e.
\begin{equation}\label{eq:risk_inExtrendedForm}
\pi(X\geq x_{\mathrm{th}}, D\geq d_\mathrm{th}| v,\tau) =
\sum_{d_k\geq d_\mathrm{th}} \pi(D = d_{k}|X\geq
x_\mathrm{th},v,\tau)\;\pi(X\geq x_{\mathrm{th}}|v,\tau),
\end{equation}
where $d_k$ is an integer denoting the number of samples
$(x^s[j],\ldots,x^{s}[j+d_k])\geq x_\mathrm{th}$,
$x^{s}[j-1]<x_\mathrm{th}$ and $x^{s}[j+d_k+1] < x_\mathrm{th}$,
which corresponds to a temporal duration of
$d_k\;\Delta$. 
\\

Based on equation (eq.~\ref{eq:risk_inExtrendedForm}), risk
(eq.~\ref{eq:RiskInAPixel}) is estimated by first calculating
probability $\pi(X\geq x_\mathrm{th}| v,\tau)$ according to
\begin{equation}\label{eq:probability1}
\pi(X\geq x_\mathrm{th}| v,\tau) =
\frac{\sum_{s=1}^S\sum_{j=0}^{N_s-1}U(x^s[j]|v,\tau,x_{\mathrm{th}})}{\sum_{s=1}^S\sum_{j=0}^{N_s-1}U(x^s[j]|v,\tau,0)},
\end{equation}
where $U$ is a step function defined as
\begin{equation}
U(x^s[j]| v,\tau,\alpha) =
\begin{cases}
            1 & \textrm{if $x^s[j]\geq \alpha$ given that $\mathrm{v}^{s}[j]\in v$ and $t_j\in\tau$}
            \\
            0 & \textrm{otherwise,}
\end{cases}
\end{equation}index $S$ is the total number of available links, $\alpha$ is the intensity threshold and $N_s$ the number
of the available scintillation samples (i.e. total number of
measurements) for link $s$ over a time window $\tau$.

Furthermore, the conditional probability $\pi(D = d_{k}|X\geq
x_\mathrm{th},v,\tau)$ is
\begin{equation}
 \pi( D=d_{k}| X\geq x_\mathrm{th}, v,\tau)  =
 \frac{\sum_{s=1}^S\sum_{j=0}^{N_s-1} H_d\left(x^{s}[j],\ldots, x^{s}[j+d_k]|v,\tau,x_\mathrm{th}\right)d_k}
 {\sum_{s=1}^S\sum_{j=0}^{N_s-1}
 U\left(x^{s}[j]|v,\tau,x_\mathrm{th})\right)},
\end{equation}
where function $H_d$ is defined as
\begin{align}
 H_d(x^{s}[j],\ldots,
x^{s}[j+d_k]|v,\tau,x_\mathrm{th}) =
 \begin{dcases}
   1\quad  \textrm{if $(x^s[j],\ldots, x^s[j+d_{k}])\geq x_\mathrm{th}$, $x^s[j+d_k+1]< x_{\mathrm{th}}$ and $x^s[j-1]< x_{\mathrm{th}}$ } \\
   \qquad \textrm{when $\{\mathrm{v}^{s}[j-1],\ldots,\mathrm{v}^s[j+d_k+1]\}\in v$ }\\
   1\quad\textrm{if $(x^s[j],\ldots, x^s[j+d_{k}])\geq
   x_\mathrm{th}$ and $x^s[j-1]< x_{\mathrm{th}}$ }\\
   \qquad\textrm{when  $\{\mathrm{v}^{s}[j-1],\ldots,\mathrm{v}^s[j+d_k]\}\in v$ and $\mathrm{v}^s[j+d_k+1]  \not\in v$}\\
 1\quad\textrm{if $(x^s[j],\ldots, x^s[j+d_{k}])\geq
   x_\mathrm{th}$ and $x^s[j+d_k+1]< x_{\mathrm{th}}$ }\\
   \qquad\textrm{when  $\{\mathrm{v}^{s}[j],\ldots,\mathrm{v}^s[j+d_k+1]\}\in v$ and $\mathrm{v}^s[j-1]  \not\in v$}\\
  1\quad\textrm{if $(x^s[j],\ldots, x^s[j+d_{k}])\geq
   x_\mathrm{th}$}\\
   \qquad\textrm{when  $\{\mathrm{v}^{s}[j],\ldots,\mathrm{v}^s[j+d_k]\}\in v$ and $\mathrm{v}^s[j-1],\mathrm{v}^s[j+d_k+1] \not\in v$}\\
   0 \quad\textrm{otherwise}
\end{dcases}
\end{align}
and gives either 1 or 0 based on the  sequence $x^{s}[j-1],\ldots,
x^{s}[j+d_k+1]$.

Finally by estimating pixel by pixel a risk
(eq.~\ref{eq:RiskInAPixel}) at each $v$, we can obtain a map in
matrix form denoted by
 \begin{equation}R_{\mathrm{ion}}^\tau \in
\mathbb{R}^{V_1\times V_2 },\;\;\;\ \mbox{where $[V_1\times V_2]$ is
the dimension of a uniform grid}. \end{equation}

\subsection{WPDOP using scintillation risks}
To estimate WPDOP (eq.~\ref{eq:WPDOP}) first we  compute matrix
$A\in \mathbb{R}^{S\times 3}$ which includes the relative positions
between a ground receiver and $S$ available satellites following the
details of appendix~\ref{Appendix:PositioningError}.

{ Regarding the weights, without further knowledge about the precise
shape of function $h()$ (the properties of which we anticipate may
be gleaned from either statistical or physical information, or
alternatively selected based on the requirements of a specific
application), we approximate $w_s$ (eq.~\ref{eq:weights}) by
\begin{equation}\label{eq:weights_new}
{w}_s:=(1-r^{(s)}_\mathrm{ion})^k,
\end{equation}
where $r^{(s)}_\mathrm{ion}$ is the ionospheric risk value at the
intersection point between the line-of-sight of (link) $s$ and the
ionospheric risk map at 350km. Weight (eq.~\ref{eq:weights_new})
satisfies the limiting values for $w_s$ (eq.~\ref{eq:weights}) as
risk varies and
parameter $k\geq 0$ 
penalizes our trust to a ray-path based on the scintillation risk
value. For the estimation of WPDOP in the result section, we set
$k=2$ to ensure that relatively high trust is given to ray-paths
that are associated with low scintillation risks.}

{So, given a hypothetical ground receiver and $S$ available
satellites, different weights are assigned to the ray-paths (between
the receiver and the satellites) based on the points the ray-paths
pierce the ionospheric map. Then, WPDOP (eq.~\ref{eq:WPDOP}) is
estimated using the weighting matrix}
\begin{equation}W =
\begin{bmatrix}
    (1-r_{\mathrm{ion}}^{(1)})^k&  \dots &0 \\
    \vdots & \ddots & \\
    0 &        & (1-r_{\mathrm{ion}}^{(S)})^k
    \end{bmatrix}.
\end{equation}
{Figure~\ref{figure:SchematicExamples}.b explains schematically how
to assigns a weight to each each ray-path using the ionospheric risk
map.}
\section{Data, map construction, results and discussion}\label{section:results}
In this section, we present ionospheric risk maps and corresponding
WPDOP ground  maps over the area of South America at geographic
latitude between $\sim -40^\circ\mathrm{N}$ and $\sim
10^\circ\mathrm{N}$ and longitude between $\sim -90^\circ\mathrm{E}$
and $\sim -35^\circ \mathrm{E}$. This region experiences significant
scintillation activity that causes deep signal fades inducing a GNSS
receiver to lose lock of one or more satellite signals and includes
the equatorial anomaly \citep{APPLETON1946}.
\subsection{$S_4$ data} Scintillation activity maximizes mostly
after sunset until a few hours after midnight local time during the
equinoctial months in the equatorial ionosphere \citep{Rezende2006}.
For this reason in this study we used available $S_4$ data between
23.00 and 03.00 Universal Time (UT). { The $S_4$ data (that was used
to produce a risk map) was measured from a network of 38 ISMR
scintillation monitors using all the available GPS, GLONNASS and
Galileo satellites over the 4 hour period. }
 In
particular, the $S_4$ measurements ( with sampling period $\Delta=1$
minute) and their corresponding \emph{IPPs} at 350km were provided
from the CIGALA/CALIBRA network - UNESP web server \citep{UNSESP}.
The \emph{IPPs} values for the interval between 23.00 and 03.00 (UT)
are depicted with black dots Figure~\ref{figuregrids}.a. The period
under investigation, which was during the night-time of 02 and 03
November 2014, was characterized by mild to strong scintillation
according to the information provided by the web software (ISMR
Query Tool) (\href{http:// is-cigala-calibra.fct.unesp.br}{http://
is-cigala-calibra.fct.unesp.br}) \citep{Vani2017}.

\subsection{Construction of ionospheric maps}
\begin{figure}[ht]
        \centering
          \includegraphics[width=0.59\columnwidth]{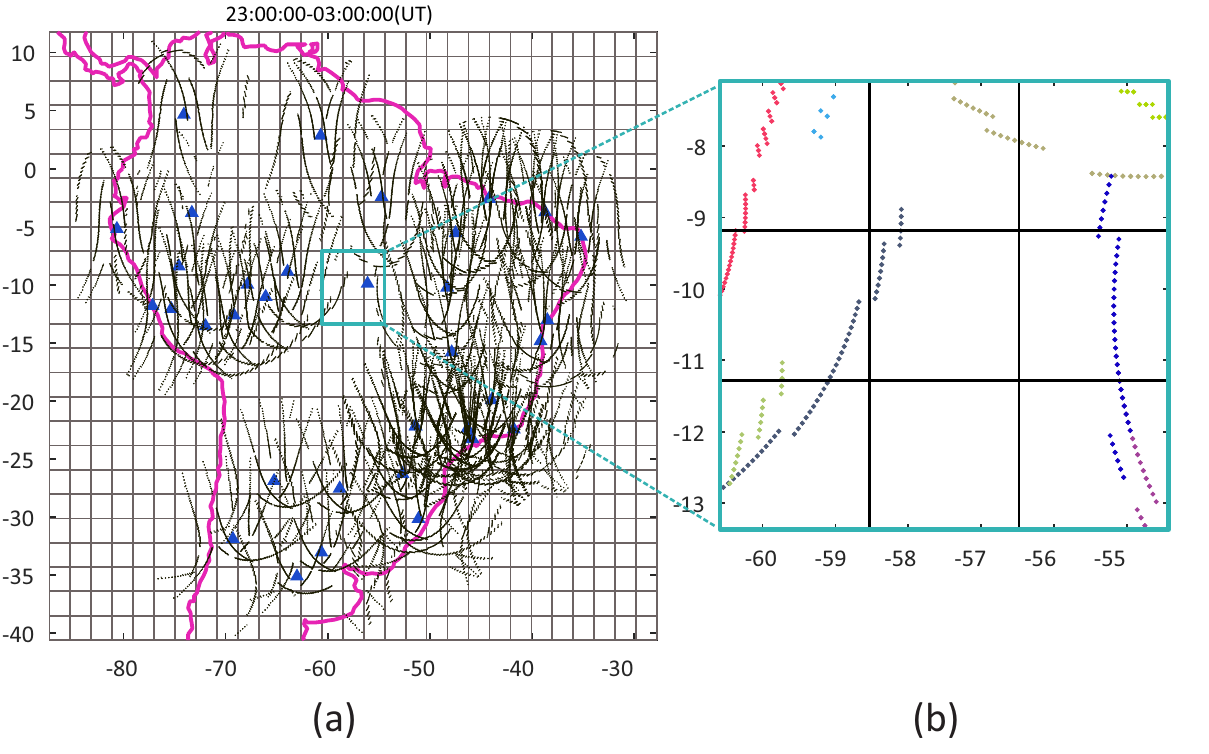}
      \caption{(a) The ionospheric thin shell at 350 km was subdivided into a grid  of spacial resolution $2^\circ\times 2^\circ$.
      The traces of all the projected $S_4$ samples (\emph{IPP data}) on the ionospheric grid at 350km over a period of 4
      hours (02 November 2014) are depicted with black dots. The locations of
      scintillation monitors are denoted by blue triangles.
      {(b)
      This figure depicts in higher resolution the square
      area marked with light blue color in Figure~\ref{figuregrids}.a.
 The traces of the projected
$S_4$ samples that  correspond to the same (satellite-scintillation
monitor) link are marked with the same color.}
      }
       \label{figuregrids}
\end{figure}

{For the ionospheric risk maps, a
uniform standard grid of spatial resolution $2^\circ\times 2^\circ$
was used for the estimation of the risk in each ionospheric pixel
(as suggested also by \citep{Takahashi2016}).  A scintillation risk
(joint probability eq.~\ref{eq:RiskInAPixel}) was estimated in each
pixel using accumulated $S_4$ data over the interval 23.00-03.00
(UT) (for example Figure~\ref{figuregrids}.b shows the available
data over the 4 hours interval for a set of nine pixels).
Ionospheric pixels that did not have any \emph{IPP data} were left
blank in the following ionospheric maps.
\\For clarity in the subsequent text, the night time interval
(23.00-03.00 UT) crossing from 02 to 03 November 2014 will be simply
referenced as that of 02 November 2014, and similarly the night time
from 03 to 04 November 2014 will be referenced as that of 03
November 2014.}
\subsection{Construction of ground maps} 
\begin{figure}[ht]
        \centering
          \includegraphics[width=1.1\columnwidth]{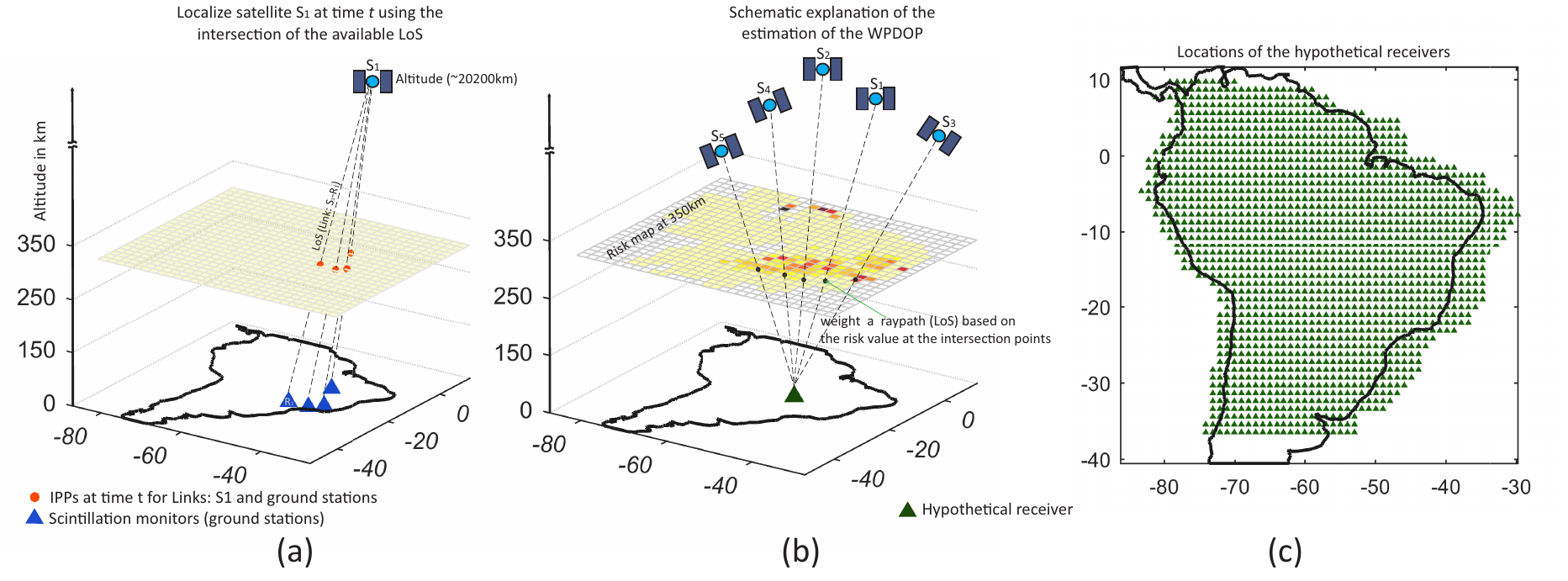}
      \caption{(a) We localize a single satellite by estimating the intersection of line-of-sights (LoS) that are formed by the ionospheric pierce points (IPPs) of $S_4$ samples
       and  the locations of the scintillation monitors (denoted by blue
      triangles) at which the $S_4$ samples were recorded.
      (b) Schematic explanation of the estimation of the WPDOP for a hypothetical ground
      receiver. In this
example we have five available satellites which intersect the risk
map at 350~km within five pixels marked with small black dots. The
weighting matrix in this example has the form
$W=\mathrm{diag}\left((1-r_{\mathrm{ion}}^{(i)})^2\right)$ where
$r_{\mathrm{ion}}^{(s)}$ are the risk values at the intersection
points between the ray-paths $s$ and the risk map at 350~km (for
$s=1,\ldots,5$). Then, matrix $W$ is inserted in
the~(eq.\ref{eq:WPDOP}).
      (c) Green triangles show the locations of hypothetical receivers on the ground (used to estimate the dilution of precision ground maps)}
      \label{figure:SchematicExamples}
\end{figure}

The spatial resolution of the
ground maps was selected as $1^\circ\times 1^\circ$ (in latitude and
longitude) by placing hypothetical receivers (denoted by green
triangles) in the centre of each pixel as shown in
Figure~\ref{figure:SchematicExamples}.c. {To calculate the WPDOP and
PDOP for a set of hypothetical ground receivers, the positions of
the satellites in view have to be estimated. There are different
ways that this can be done using for example orbital information. In
our study, the locations of the available satellites were computed
geometrically at time $t$ using downloaded data from \cite{UNSESP}.
Particularly, the downloaded data provided, in addition to $S_4$
measurements, the IPP geographic coordinates at 350km (between the
scintillation monitor and the satellite), the time of data
acquisition and the satellite-monitor information (i.e. the PRN code
of the satellite and name/location of the scintillation monitor) for
each acquired $S_4$ measurement. Thus, the IPPs and locations of
scintillation monitors that are associated with the same satellite
at time $t$ were used to form 3D lines. The extensions of these
lines were intersected at the location of the satellite (as depicted
in
Figure~\ref{figure:SchematicExamples}.a).} 

{ Furthermore, given the positions of the satellites at time $t$
over South America, the satellites in view for a hypothetical ground
receiver were determined based on an elevation angle cut-off of
$20^\circ$. Then, matrix $A$ in (eq.~\ref{eq:WPDOP}) was calculated
using the locations of the satellites in view and the hypothetical
receiver. For each (satellite-hypothetical receiver) line-of-sight
(LoS), each weight (eq.~\ref{eq:weights_new}) in the diagonal matrix
$W$ was computed using the risk value at the pixel where the line
pierced the ionospheric map at 350km.
Figure~\ref{figure:SchematicExamples}.b illustrates an example of
the estimation of the WPDOP for a hypothetical ground receiver when
the ionospheric risk map is taken into account and five satellites
are in view. }

\subsection{Ionospheric risk maps over South
America}\label{sec:MapsIon} 
\begin{figure}[ht]
     \centering
      \centering
          \includegraphics[width=1\columnwidth]{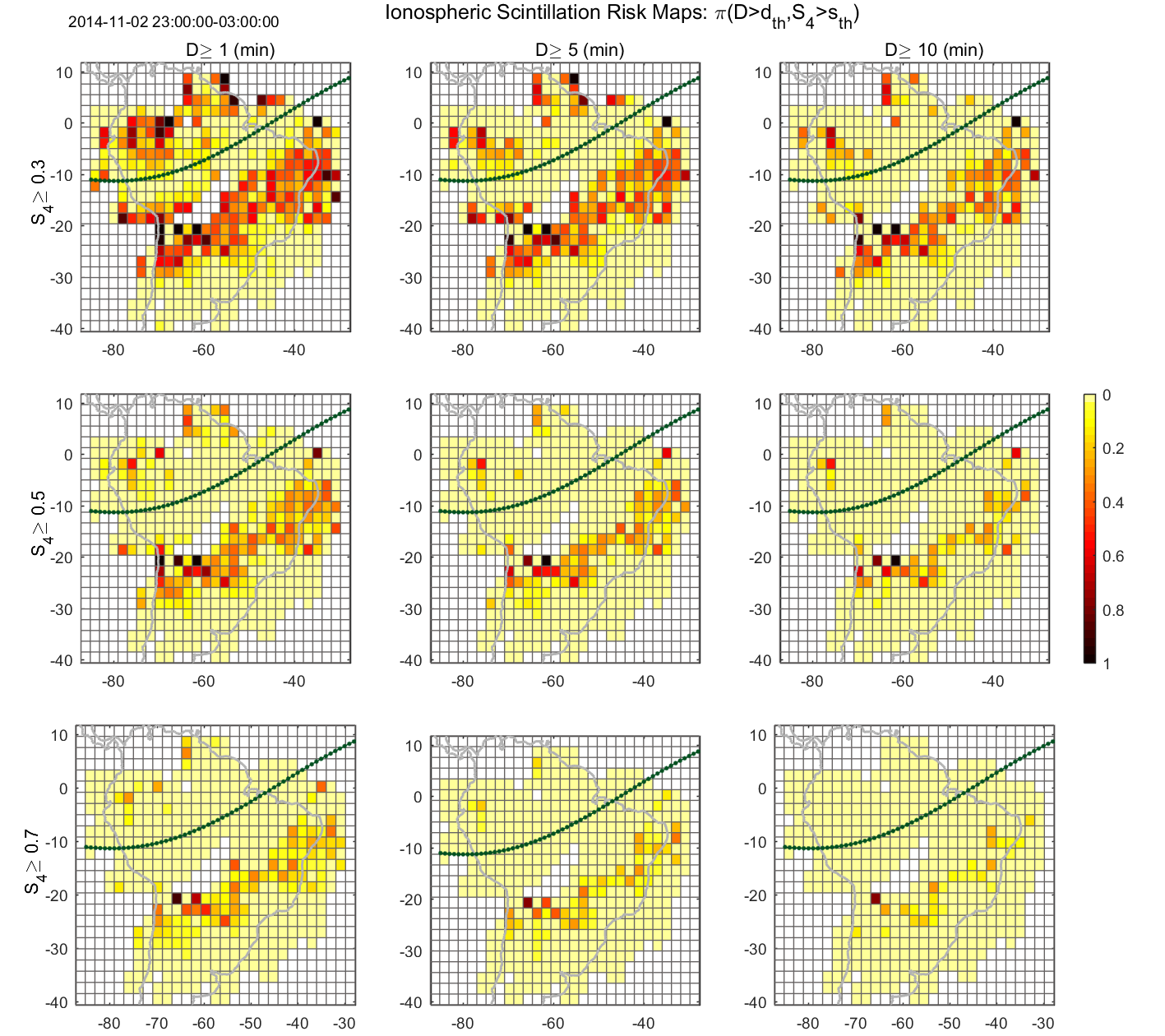}
      \caption{Ionospheric risk maps for different $S_4$ thresholds and durations for 02 November 2014 (rows correspond to $S_4$ threshold 0.3, 0.5 and 0.7, and columns correspond to duration thresholds of 1, 5, 10 minutes).
       The risk maps (based on the joint
probability eq.~\ref{eq:RiskInAPixel}) were estimated using
accumulated $S_4$ over the interval 23.00-03.00 (UT). The dash green
line depicts the geomagnetic equator.}
 \label{figure_day2}
\end{figure}

\begin{figure}[ht]
     \centering
      \centering
          \includegraphics[width=1\columnwidth]{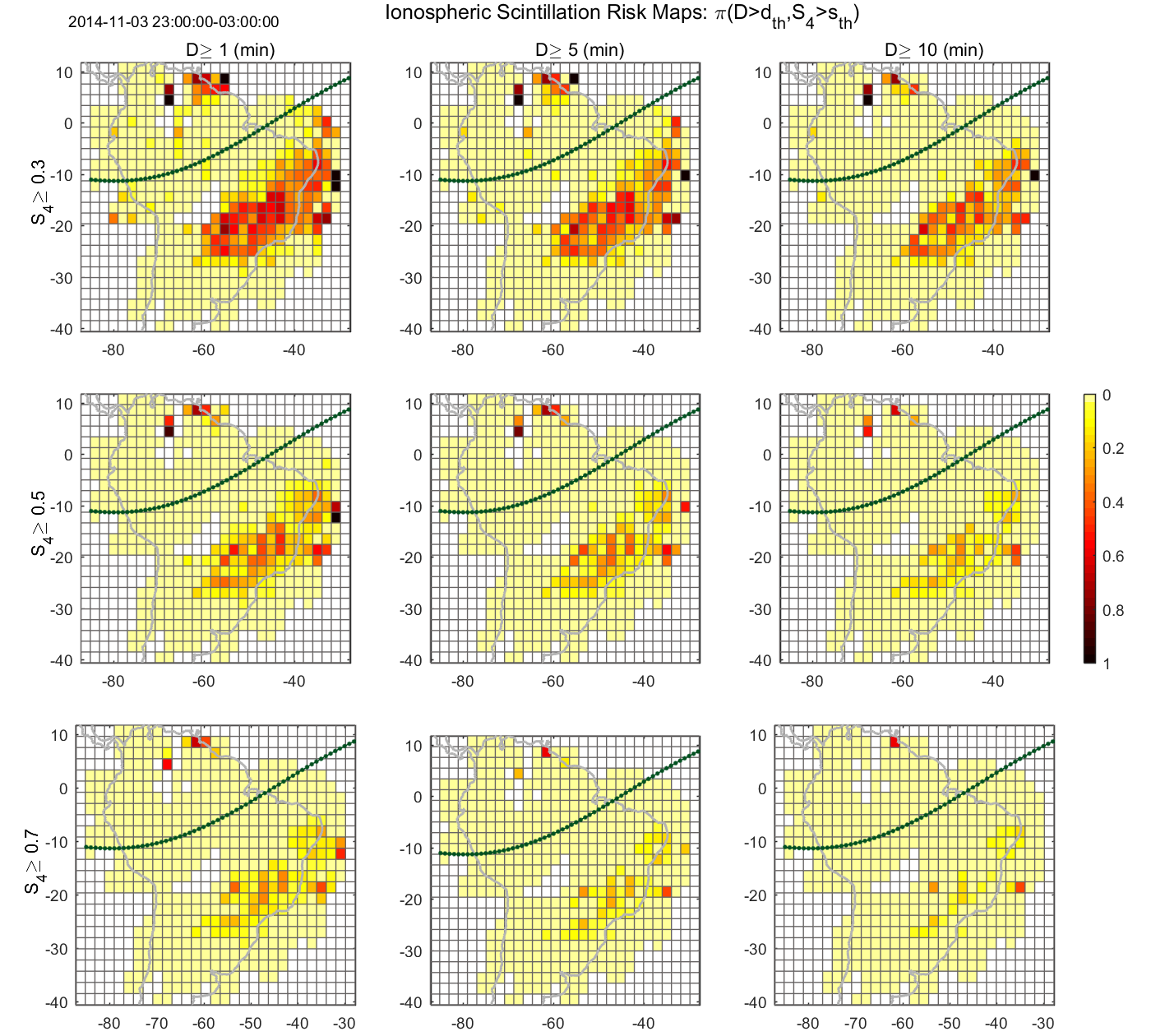}
      \caption{Ionospheric risk maps for different $S_4$ thresholds and durations for 03 November 2014 (rows correspond to $S_4$ thresholds 0.3, 0.5 and 0.7,
      and columns correspond to duration thresholds of 1, 5, 10 minutes). The risk maps (based on the joint
probability eq.~\ref{eq:RiskInAPixel}) were estimated using
accumulated $S_4$ over the interval 23.00-03.00 (UT)}
 \label{figure_day3}
\end{figure}

Figure~\ref{figure_day2}
and~\ref{figure_day3} show the ionospheric risk maps based on the
joint probability of scintillation intensity and duration, using
different thresholds for the $S_4$ value and duration of the
\emph{scintillation event}. More precisely, in
Figures~\ref{figure_day2} and ~\ref{figure_day3} the rows correspond
to $S_4$ thresholds 0.3, 0.5 and 0.7 respectively, and the columns
correspond to duration thresholds of 1, 5, 10 minutes respectively.
The green line designates the geomagnetic equator and the white
pixels correspond to ionospheric regions where the risk was not
estimated due to lack of data (the ionospheric regions where
\emph{IPP data} was available appears in the left image of
Figure~\ref{figuregrids} for 02 November 2014). As expected, based
on the results of Figure~\ref{figure_day2} we can observe that as
the thresholds values are getting higher the risk decreases.

Moreover, we can see that there are two bands of scintillation i.e.
north and south of the geomagnetic equator. This corresponds to the
well known equatorial anomaly \citep{Kintner2007a}. The risk of
scintillation in the north band seems to disappear for increasing
thresholds relatively drastically (see second and third row in
Figure~\ref{figure_day2}) which indicates that scintillation of
shorter duration and lower intensity prevails in this ionospheric
region. We note that this could also be related to the number of the
ground scintillation monitors which are less in the north band
compared to the south band (so the south band may give us more
reliable estimates).

In Figure~\ref{figure_day3} scintillation structures remain similar
as in Figure~\ref{figure_day2}, however, we can notice that the
scintillation in Figure~\ref{figure_day2} covers a broader area of
the south band compared to Figure~\ref{figure_day3} where high
scintillation activity is observed predominantly over Brazil.
Overall, we can say that scintillation of low intensity ($0.3 \leq
S_4 < 0.5$) affects a broad area of South America with risks
approaching 0.8 whereas scintillation risk drops drastically for
$S_4\geq 0.7$ during both days.

\FloatBarrier

\subsection{Ground maps over South America} 
\begin{figure}[ht]
 \centering
 \begin{tabular}{c}
           \includegraphics[width=0.98\columnwidth]{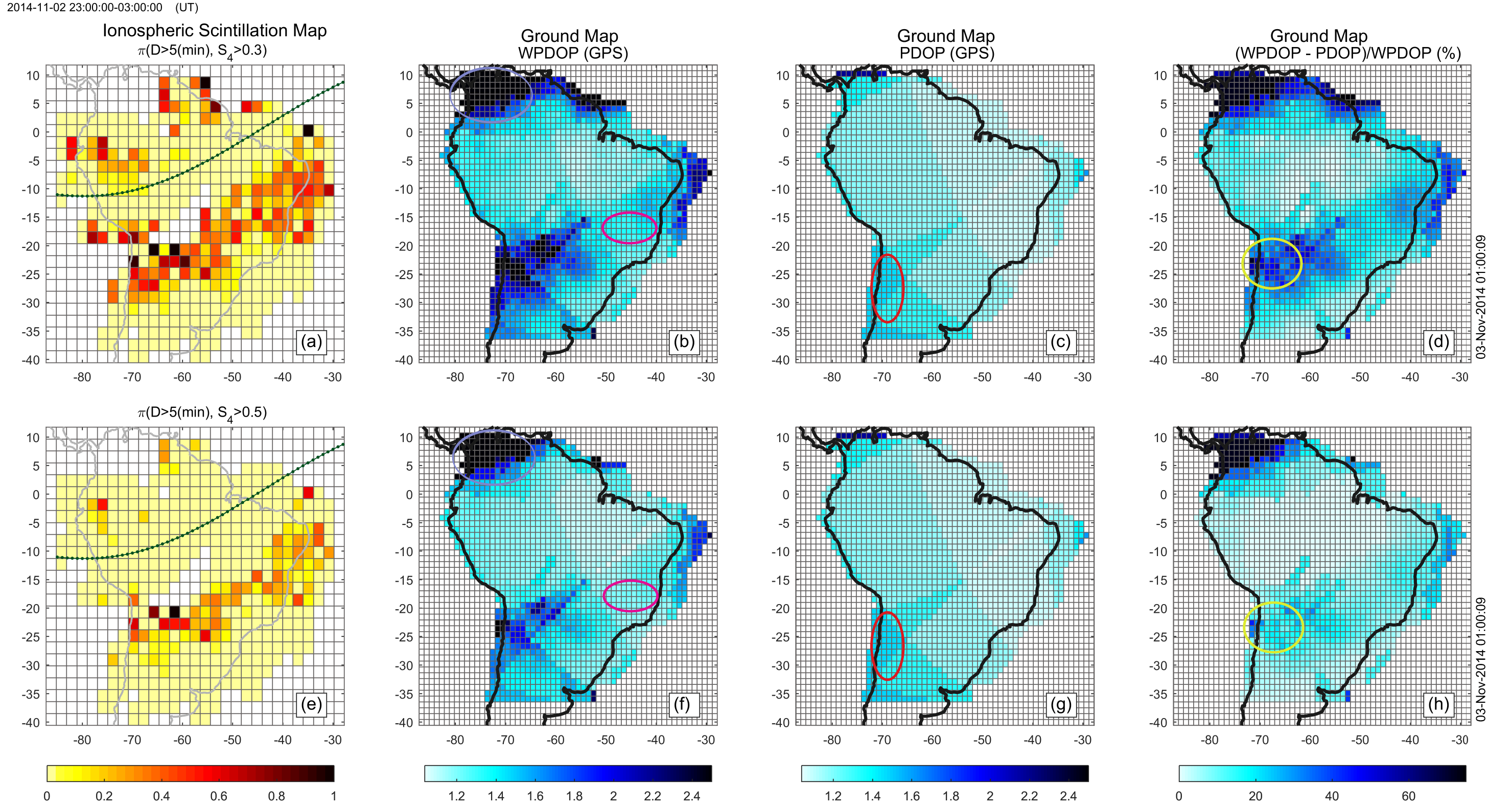}
            \\ (i) Ionospheric risk map estimated using $S_4$ data over 23.00-03.00 (UT)\\ and ground maps using only GPS constellation at 01.00.09 (UT)\\
            \includegraphics[width=0.98\columnwidth]{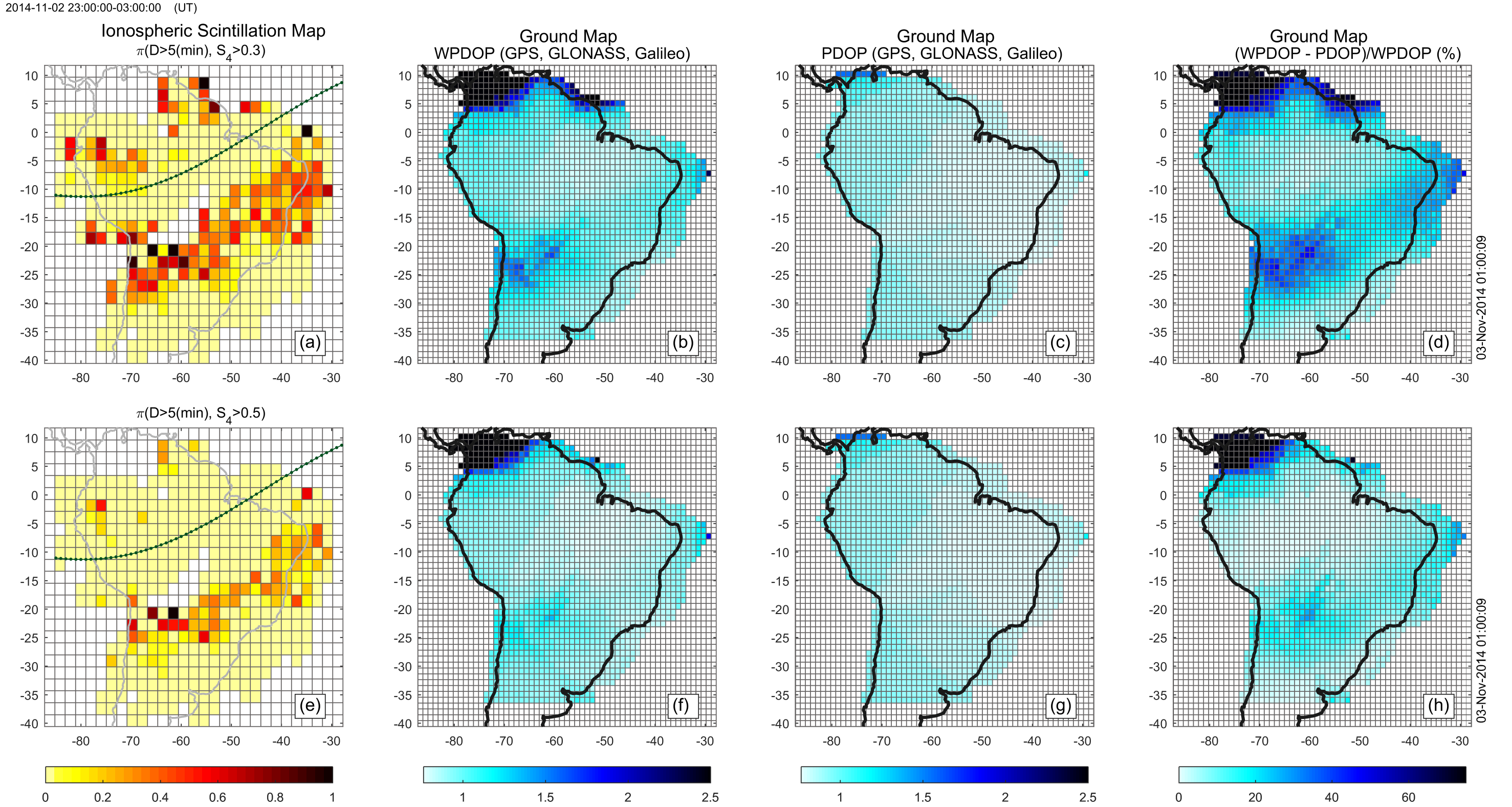}
            \\ (ii) Ionospheric risk map using $S_4$ data over 23.00-03.00 (UT) \\ and ground maps using GPS, GLONASS and
            Galileo constellation at 01.00.09 (UT)
 \end{tabular}
 \caption{Risk maps and ground maps for different sets of satellite constellations for
 02 November 2014. Ionospheric static risk maps were estimated using data over
 the interval 23.00-03.00 (UT). The geometry of the constellations
 for the ground maps
 was that of time instant 01.00.09 (UT).
 Within (i) or (ii) each row corresponds to different thresholds in
 $S_4$.  From left to right, column 1 shows ionospheric risk maps for thresholds in $S_4$ of 0.3 and 0.5, and
 scintillation duration of 5 minutes, column 2 illustrates WPDOP ground maps which explicitly incorporate the scintillation risks as weights from ionospheric map in column
 1, column 3 gives the PDOP ground maps (with no scintillation being included), column 4 give us the difference between WPDOP and PDOP, expressed as a percentage of WPDOP.
 The regions that are described in the text are marked with colored ellipses.}%
 \label{figure_grday2}
\end{figure}

We illustrate the
effect of scintillation on the ground by estimating the proposed
weighted dilution of precision in a set of hypothetical ground
receivers (small green triangles in
Figure~\ref{figure:SchematicExamples}.c) using the risk maps
estimated in the previous section and different sets of satellite
constellations. Two representative examples of instant ground maps
are given in Figure~\ref{figure_grday2} and~\ref{figure_grday3} for
02 November 2014 and 03 November 2014 respectively. For each day,
the WPDOP and PDOP ground maps were calculated by utilizing the
geometry of satellite constellations at precisely 01.00.09 (UT). The
weights for the WPDOP maps were based on the ionospheric risks maps
estimated in subsection~\ref{sec:MapsIon} using $S_4$ data collected
over 23.00-03.00 (UT). {In particular, a weight
(eq.~\ref{eq:weights_new}) for a link was computed using the risk
value at the point where a receiver-satellite ray-path pierces the
ionospheric risk map (at 350 km).} The risk maps (a) and (e) were
calculated considering duration threshold of 5 minutes and $S_4$
intensity threshold either 0.3 or 0.5 (as shown in Figure
~\ref{figure_grday2} and~\ref{figure_grday3}). Two sets of ground
map estimations were computed to show the effects of scintillation
with respect to the number of available constellations: (i) GPS (7
satellites) and (ii) GPS, GLONASS and Galileo (17 satellites). The
elevation angle cut-off was set to $20^\circ$, i.e. ray-paths  which
were lower than this threshold were discarded from the PDOP and
WPODP calculations.

In Figure~\ref{figure_grday2} (and similarly in
Figure~\ref{figure_grday3}), along column 1 we can observe the
ionospheric risk maps; along column 2 the maps of WPDOP estimated
based on the risk maps (of column 1); along column 3 the map of PDOP
(we note that maps (c) and (g) are identical and have been repeated
to ease the comparisons); and along column 4 the contribution of
scintillation to the total dilution of precision (in a percentage
form).
{In Figure~\ref{figure_grday2}.i, PDOP values in the ground map (c)
and (g) are low and increase slightly towards the west coast;
in  particular in the area marked with the red ellipse
the PDOP values are higher than in the corresponding area in
Figure~\ref{figure_grday2}.ii (map (c) and (g)) due to less
satellite coverage.
In general, WPDOP values (maps (b) and (f) along column 2) in
Figure~\ref{figure_grday2}.i are increased mainly in the region
under the scintillation structures. However, for longitude between
$\sim -50^\circ$E and $\sim -40^\circ$E and latitude between $\sim
-20^\circ$N and $\sim -15^\circ$N, (i.e. the area marked with a pink
ellipse in maps (b) and (f)) the WPDOP values are relatively low
even though the risk due to scintillation over that region is
significant. This is most likely because the ray-paths of the
satellites in view (that were used to estimate the WPDOP at 01.00.09
(UT)) do not intersect the scintillation active region at this
particular moment. Moreover, in the Northwest of the continent (area
marked with blue circles in maps (b) and (f) of
Figure~\ref{figure_grday2}.i), we can observe high WPDOP values. The
higher values of WPDOP in that area (compared to the PDOP values in
map (c) of Figure~\ref{figure_grday2}.i in the same area) indicate
that $w_s\ll 1$. Low weights mean that the ray-paths between the
hypothetical receivers located in the Northwest and the satellites
in view  at 01.00.09 (UT) pierce the ionospheric map in pixels with
high risk values based on (eq.~\ref{eq:weights_new}).

WPDOP ground map (f) in Figure~\ref{figure_grday2}.i (column 2 and
second row) reveals that high $S_4$ threshold (i.e $S_4\geq0.5$)
reduces the effect of scintillation on the WPDOP values. This is
because the strong scintillation activity is more focal based on the
corresponding risk map (e) in Figure~\ref{figure_grday2}.i. However,
by comparing the percentage maps (d) and (h) along the column 4 of
Figure~\ref{figure_grday2}.i, we can observe that the loss in
dilution of precision is localized similarly both when $S_4\geq 0.3$
and $S_4\geq 0.5$. Both (WPDOP-PDOP)/WPDOP maps of column 4 (in
Figure \ref{figure_grday2}.i) give a percentage of at least $30\%$
contribution from scintillation which can be observed for longitude
between $\sim -60^\circ$E and $\sim -75^\circ$E and latitude $\sim
-20^\circ$N and $\sim -25^\circ$N (area marked with yellow ellipses
in map (d) and (h)). This suggests that scintillation has a
significant influence in the position accuracy and has to be taken
into account. Furthermore, we can notice that further away from the
ionospheric scintillation structures, the scintillation contribution
to WPDOP drops to zero.

With the introduction of extra constellations, the effect of
scintillation is reduced broadly with a small exception in the north
part of the continent as we can observe in WPDOP maps (b) and (f)
along column 2 of Figure~\ref{figure_grday2}.ii (area marked with an
ellipse in red color). This happens due to scintillation and not due
to poor geometry since the PDOP values of the PDOP map (c) and (g)
in column 3 of Figure~\ref{figure_grday2}.ii. appear to be small in
that region. In general, we can see along columns 4 of
Figure~\ref{figure_grday2}.i and \ref{figure_grday2}.ii that the
percentage contributions from scintillation, maps (d) and (h), are
broadly similar which implies that, while increasing the number of
constellations (as currently shown) decreases the WPDOP values, the
effect of scintillation is not totally eliminated.

Furthermore, test case of Figure~\ref{figure_grday3} demonstrates
that the effect of scintillation activity can yield different ground
structures. This is happening because the geometry of the available
constellations and the risk maps are different between
Figure~\ref{figure_grday2} and Figure~\ref{figure_grday3}. The maps
(d) and (h) along column 4 in Figure~\ref{figure_grday3} reveal
``shadowing effects'' on the Earth due to scintillation around the
ionospheric active region, with the contribution decreasing with
increasing distance from the scintillation structures. Similarly as
in the example of Figure~\ref{figure_grday3} we can conclude that
the effect of scintillation on the ground is reduced when more
satellites are in use and when the high risk regions in the
ionospheric map are more localized.
 \begin{figure}[htb]
 \centering
 \begin{tabular}{c}
           \includegraphics[width=0.98\columnwidth]{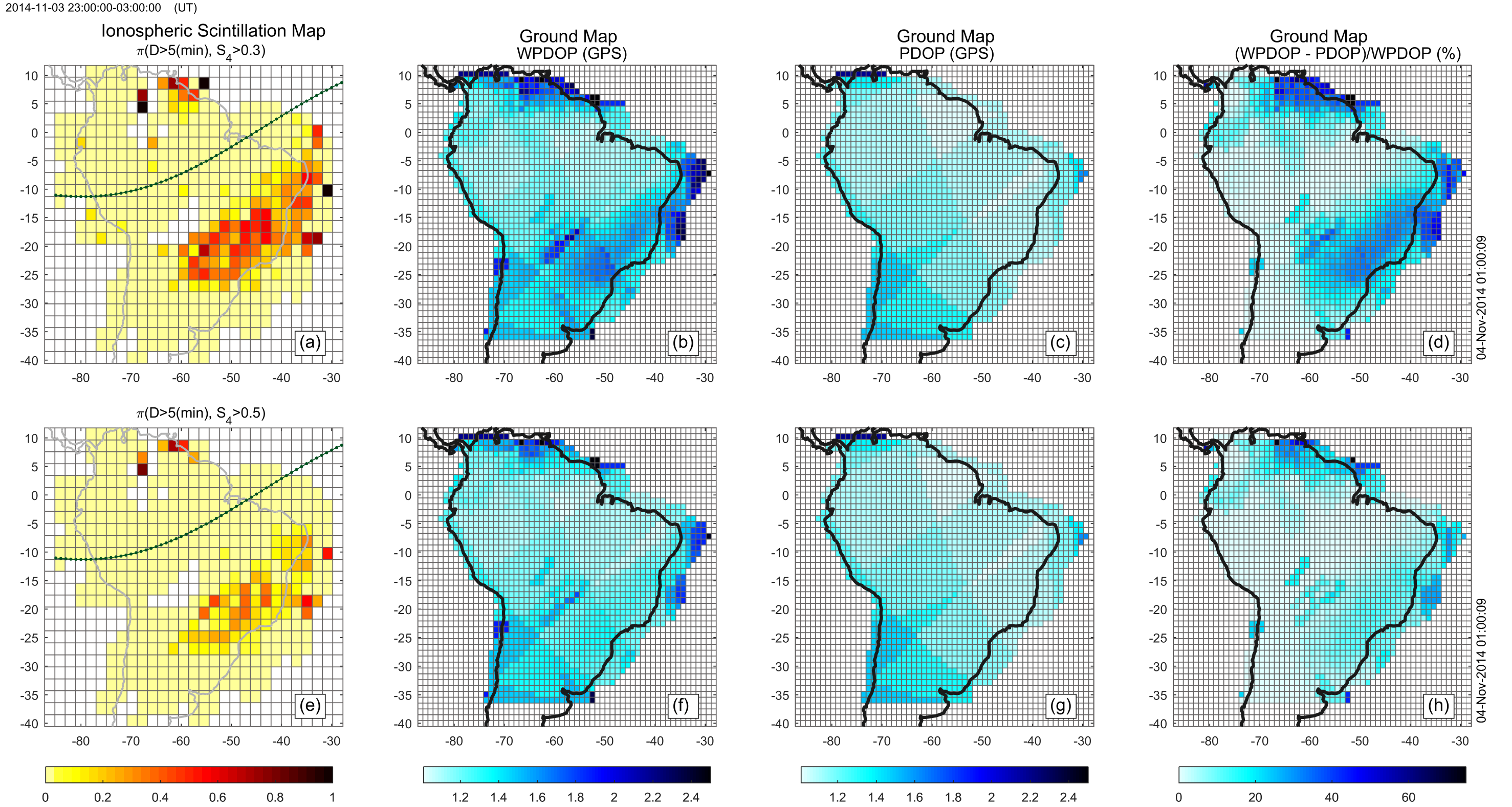}
            \\ (i) Ionospheric risk map estimated using $S_4$ data over 23.00-03.00 (UT)\\ and ground maps using only GPS constellation at precisely 01.00.09 (UT)\\
            \includegraphics[width=0.98\columnwidth]{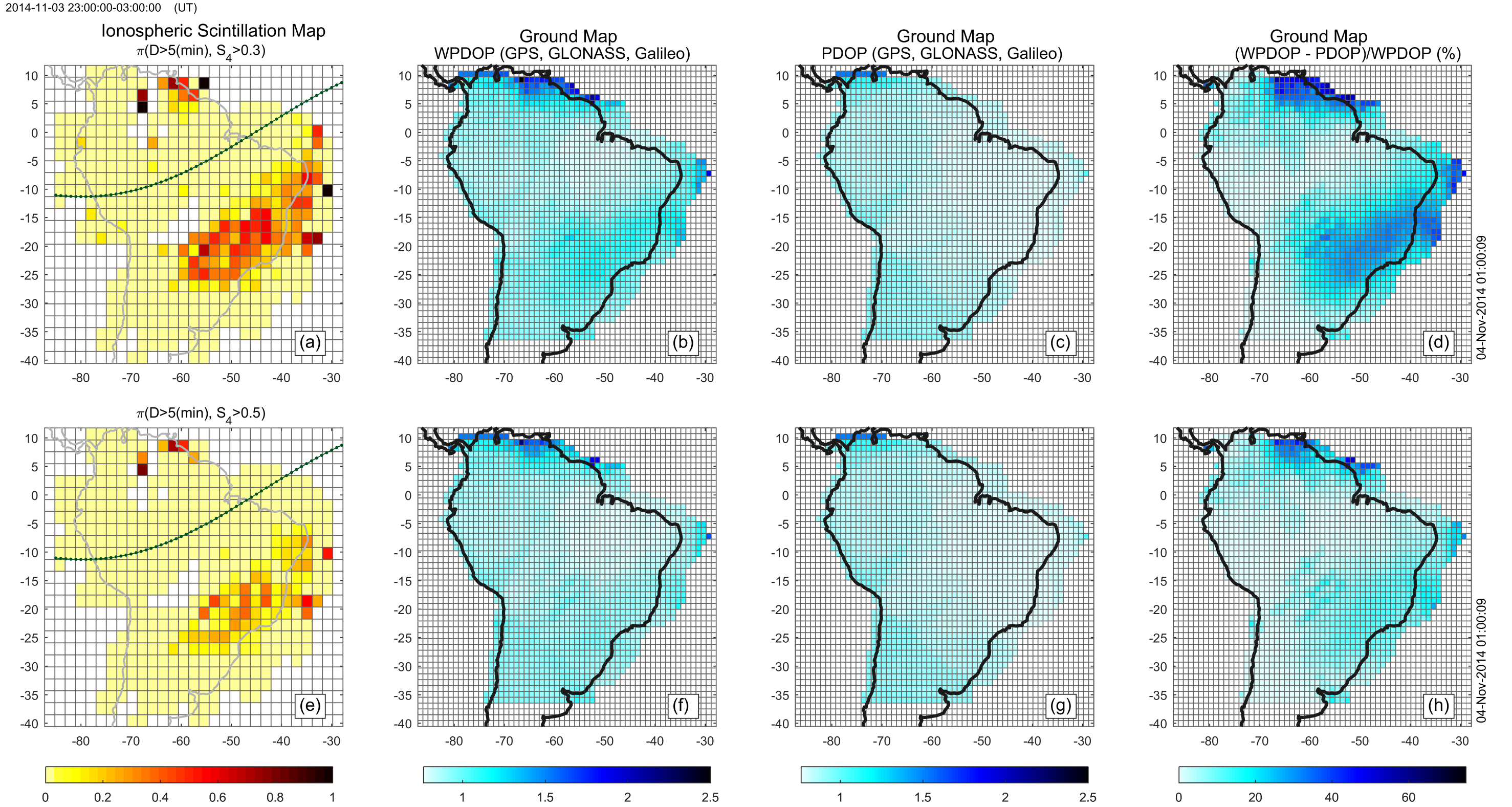}
            \\ (ii) Ionospheric risk map using $S_4$ data over 23.00-03.00 (UT)\\ and ground maps using GPS, GLONASS and
            Galileo constellation at 01.00.09 (UT)
 \end{tabular}
 \caption{The results are presented as in Figure~\ref{figure_grday2}. Ionospheric risk maps and ground maps for different sets of satellite constellations for 03 November 2014.}
 \label{figure_grday3}
\end{figure}

\FloatBarrier
\subsection{Discussion}\label{sec:Discussion}
\subsubsection{Current implementation and intrinsic limitations}
{In the current application of our methodology, we were restricted
by the  number of available ground scintillation monitors
(unfortunately, at the moment the ground monitors are very few and
unevenly spread over the continent) and the fixed number of
satellites. So, in effect these limitations forced us to produce
ionospheric scintillation risk maps using data collected over a
relatively long time window. The selection of the 4 hour time window
(period of the highest scintillation activity) was used to capture
the average scintillation activity. Therefore, the estimated static
risk maps described the average scintillation activity during the
specific time window.}

{A WPDOP value was estimated for a specific locations using the
number of available satellites at that moment and the risk map, as
shown in Figure~3.b. The weights were
 assigned to different ray-paths based on the  ionospheric scintillation
risk maps and were directly related to statistical information about
scintillation since the risks that were used to estimate the weights
(eq.~26) had been computed from scintillation data (i.e.
scintillation intensity and duration which has been extracted from
$S_4$ measurements). Therefore, the proposed WPDOP included the
impact of the average scintillation activity.}

{ In the future, if more scintillation monitors are built on the
ground then our methodology can produce ionospheric scintillation
maps with higher temporal resolution (say, every 30 or 15 minutes).
This will allow to update the weights in the WPDOP constantly and to
observe the impact of the dynamic scintillation activity. Currently,
we can say that a moving window and/or a shorter time window for the
data collection (to construct ionospheric risk maps) could be
implemented only in areas that are covered by a dense network of
scintillation monitors. By using shorter data collection intervals
and considering for example lower scintillation thresholds, we could
produce scintillation risk maps useful in aircraft navigation. These
risk maps can update the weights in the PDOP and provide information
to the pilot whether to trust the navigation system.}
\subsubsection{Comparing WPDOP with PDOP}
The dilution of precision is used as a quality measure in
positioning systems since it expresses the expected uncertainty in
the positioning estimates. Roughly speaking, a low dilution of
precision implies high confidence level in the positioning results,
and a high dilution of precision indicates low confidence level.

In general, we can notice that the standard PDOP considers
implicitly the scintillation effects in the sense that strong
scintillation can lead to loss of one or more satellites resulting
in high measured PDOP values (due to poor satellite geometry of the
remaining satellites). However, by using the standard PDOP it is not
possible to quantify the impact that low-to-high scintillation
activity has on the accuracy of a positioning system when losses of
lock do not occur. In such cases, the degradation of the GNSS
positioning may be significant even though the satellite
constellation geometry would appear promising. The concept of WPDOP
allows to estimate the impact that scintillation induced errors can
have on PDOP by assigning the proposed scintillation weights in all
the available satellite-receiver links.

Overall, we observed that the more satellites were used in the
estimation of the dilution of precision, {the lower were the values
both for PDOP and WPDOP and hence lower position errors were
expected}\footnote{For further details about the link between
WPDO/PDOP and position error see
Appendix~\ref{sec:PDOPandPositionError}}. So, the combination of
GPS, GLONASS and Galileo can improve precision. However, there were
still some areas which were affected by scintillation (especially
for weak scintillation of duration 5 minutes or more) even when all
the available constellations were used. This suggests that by
increasing the number of satellites in view, the positioning
precision can be further improved in the case of strong localized
scintillation. This outcome is in line with a recent work by
\citep{Marques2018} and the review article \citep{Langley2012}.

\subsubsection{Novelty and transferability}

In this work, instead of using basic statistical analysis, we
exploited the concept of risk from decision theory to quantify the
effects that the ionospheric scintillation can have on potential
end-user
activities. 
{In particular, we derived risks considering that the quality of a
satellite-receiver communication can be characterized only by
scintillation measurements and then we constructed ionospheric risk
maps using $S_4$ data provided by the UNESP server in South
America.} As we showed in section~\ref{sec:MapsIon} this methodology
can be used for the construction of sequences of ionospheric risk
maps (Figure.~\ref{figure_day2}and~\ref{figure_day3}). Such maps
could be further used as a-priori information for ionospheric
stochastic models, for enriching scintillation climatology
data-bases, or for improving understanding of scintillation
phenomena over multiple years \citep{Priyadarshi2015,Jiao2018a}.
Moreover, by explicitly incorporating scintillation risks into
dilution of precision, one can  overcome intrinsic  limitations of
empirical models and receiver specifications. Finally, the proposed
methodology allows to produce high resolution ground map which could
potentially be used to help mitigate the effects of ionospheric
scintillation on GNSS positioning.

The proposed methodology for the construction of ionospheric risk
and ground maps is reusable and transferable. Notably, if one wants
to tailor risk to specific applications, then this methodology has
the advantage that its parameters can be optimized and validated for
specific applications. For example the loss function
(eq.~\ref{eq:binaryLoss_new}) allows us to select scintillation
intensity and duration thresholds, $x_\mathrm{th}$ and
$d_\mathrm{th}$ respectively, based on the specifications of the
receiver and the needs of safety critical applications. By employing
a 0-1 loss function, we showed that risks can be easily interpreted
in terms of probability of \emph{scintillation events}. However, not
only the thresholds, but also alternative loss functions to
equation~(eq.~\ref{eq:binaryLoss_new}) can be adopted. Moreover, our
risk formalism allows to consider extra (measurable) inputs in
 feature $Z$ (of section \ref{sec:preliminary}) in addition to scintillation data $X$ and durations of events $D$, such as
information about data gaps and other ionospheric quantities (e.g.
total electron content (TEC)). Therefore, the risks
(eq.\ref{eq:Risk})
(as joint probabilities of multiple parameters) can be estimated 
to illustrate an overall ionospheric activity.
{Regarding  the
weighting factors $w_s$ (eq.~\ref{eq:weights}), different shapes of
the error variance function $h(.)$ (eq.~\ref{eq:function_h}) can be
adopted based on the outcomes of future research without any changes
in the proposed methodology of section~\ref{sec:WPDOPtheory}}.

 Alternatively, the construction of multiple risk maps using different sets of
 inputs (measurable features),
 such as scintillation
 values measured in different frequencies
which reflect how small-scale plasma density irregularities in F
region affect different propagating signals and  gradients of the
TEC which can capture large scale irregularities, can offer
versatile and complimentary information to the users. As a further
step, these sets of risk maps can infer implicit statistical
information about different (multi-frequency) errors which can be
considered in the position error model (eq.~\ref{eq:linearSystem})
and the estimation of the weights in the WPDOP. Hence, the concept
of WPDOP can be extended to include other ionospheric-related errors
to inform how different propagation errors arising from ionospheric
structures can impact the positioning on the ground.

In particular, our formalism enables the adaptation of the WPDOP
given the presence of scintillation or any other ionospheric-related
errors (e.g. in the equatorial  F-region). For example the error
term in the observation model (eq.~\ref{eq:linearSystem}) could be
decomposed into more uncorrelated error terms e.g. $\varepsilon =
\varepsilon_\mathrm{sc}+\varepsilon_{\mathrm{\Delta
TEC}}+\varepsilon_{\mathrm{rem}}$ and the corresponding ionospheric
risks could be used to provide information about the different error
variances (expressed as weights) for the WPDOP, e.g.
$$w_s=\frac{\gamma}{\gamma_{\varepsilon^{(s)}}}=\frac{\gamma}{\gamma+
h(r_\mathrm{sc}^{(s)})+g(r_{\mathrm{\Delta TEC}}^{(s)})}
\quad\textrm{ for}\quad s=1\dots S,
$$
where $g(r_{\mathrm{\Delta TEC}}^{(s)})$ is the error covariance due
to $\Delta TEC$ ionospheric effects. Therefore, based on the
available statistical knowledge, a WPDOP calculated by any
hypothetical receiver could depend upon the combination of
scintillation levels on multiple frequencies (utilized for
positioning), the presence of losses of lock on some or all the
frequencies associated with a given link, other ionospheric errors
(e.g. TEC) and the geometry of the available links. Then by
comparing WPDOP against PDOP, we can infer information about the
impact that different ionospheric irregularities can have on the
positioning accuracy on the ground.


\section{Conclusions}\label{sec:conclusions}
This paper proposed a methodology which connects the ionospheric
uncertainties in the sky originating from small-scale irregularities
in the F-region (e.g. in the equatorial ionosphere) with their
impact on the positioning on the ground. In particular, first we
exploited decision theory to develop risks defined as expected
losses that incur during a satellite communication activity which is
assumed to be fully characterized by ionospheric measurable
features, in our case scintillation intensity and duration. The
proposed scintillation risks  relied on a 0-1 loss function and thus
they were equal to the joint probability of scintillation intensity
and duration
above specified thresholds. 
 The derived risk formulation was used to estimate
pixel-wise risk values in an uniform ionospheric grid at 350 km
and the estimated risk maps reflected the ionospheric plasma density
inhomogeneities causing scintillation during the observed time
window.

We demonstrated our methodology by estimating risk maps in the sky
using accumulated data obtained from GPS, GLONASS and Galileo
scintillation monitors in South America during local night-time
hours and using different scintillation intensity and duration
thresholds. Overall, we observed that the two bands of scintillation
i.e. north and south of the geomagnetic equator were visible in
cases where the intensity threshold did not exceed value 0.5 and
when the duration of the {\it scintillation events} was up to 10
minutes. Subsequently, to understand the effect of scintillation on
the ground, we proposed to use a weighted position dilution of
precision (WPDOP) which was estimated by assigning different weights
to receiver measurements from different satellites using the
ionospheric scintillation risks along the line-of-sights (LoSs) of
the corresponding links. Thus, the dilution of precision on the
ground depended upon the combination of the geometry of the
available links and the scintillation-induced observable errors
(quantified though the scintillation risks) on the same available
links.

Finally, we constructed instantaneous WPDOP ground maps combining
information from the ionospheric risk maps and the available link
geometries. The WPDOP maps revealed those areas on the ground which
were more affected by ionospheric scintillation. We noted that the
scintillation effects on the ground receiver were more prominent
when only GPS was available. However, the effects were not totally
eliminated with the addition of extra constellations.

The present work focused on the effect of ionospheric scintillation
to the dilution of precision. In the future, other sources of errors
or modelling uncertainties can be taken into account in the dilution
of precision estimations (i.e. superposition of risks associated,
not only with scintillation but also with other ionospheric
measurable features, can be considered). Also, we expect to further
study how errors due to scintillation limit the accuracy in position
estimation both from theoretical and applied perspective and how
statistical information about these errors can be incorporated into
existing precise point
positioning softwares. 

\section*{Author contribution statement}
Methodology for ionospheric maps: AK, NDS and BCV with  support from
the other co-authors; Theory for ground maps: AK, NDS  and VR;
Implementation of ionospheric maps: AK and BCV; Implementation of
ground maps: AK; Analyzing the results: AK and NDS; Writing the
paper: AK as main author in collaboration with NDS and VR and with
editorial support from BCV, IA and BF; Scoping the research and
ongoing discussion: BF and IA.

\section*{Data availability statement}
The $S_4$ data used to estimate the ionospheric maps in
section~\ref{sec:MapsIon} was downloaded from
\href{http://is-cigala-calibra.fct.unesp.br/is/ismrtool/map/service_getMapIppPoints.php?date_begin=2014-11-01 00:00:00&date_end=2014-11-04 24:00:00&satellite=GPS,GLONASS,GALILEO,SBAS&station=24,24,25,11,37,1,15,28,2,3,4,5,6,14,7,8,29,26,46,47,48,49,50,51,52,53,54,55,56,57,58,59,61,62,63,64,66,67,68,69,70,71,72,73,74,75,76,77,78,79,80,81,82,83,84,85,86,87,45,33,44,41,32,30,31,34,43,42&filters=elev%3E=20;&field=s4&aggregation=none&ion=350&mode=csv
}{http:// is-cigala-calibra.fct.unesp.br}; The
MATLABa\textregistered codes for the estimation of the ionospheric
and ground maps can be downloaded from the link\\
\href{https://github.com/AlexandraPouk/Ionospheric-Scintillation-Maps-and-PDOP/tree/master}{https://github.com/AlexandraKoulouri/Ionospheric-Scintillation-Maps-and-PDOP}.
.

\section*{acknowledgements}
This work was supported by the Natural Environment Research Council
(NERC), UK, (NE/R009082/1). AK was supported by the Academy of
Finland Postdoctoral Researcher program (no 316542). BV thanks
Federal Institute of Education, Science and Technology of Sao Paulo
(IFSP), Sao Paulo State University (UNESP) and CAPES (CAPES/PDSE n.
19-2016/Process no. 88881.134266/2016-01) for supporting his
research. The authors would like to thank Prof. J. F. G. Monico,
Department of Cartography, Sao Paulo State University (UNESP) and
Prof. M. H. Shimabukuro, Department of Computer Science, Sao Paulo
State University (UNESP) for providing the $S_4 (\mathrm{L}1)$  data
that was deployed in the context of the Projects CIGALA and CALIBRA
(both funded by the European Commission (EC) in the framework of the
FP7-GALILEO-2009-GSA and FP7GALILEO2011GSA1a, respectively), and
FAPESP Project Number 06/04008-2. The ground stations and $S_4$
data, that were used to demonstrate the proposed methodology, are
currently maintained by the National Institute of Science and
Technology - GNSS Technology to Support Air Navigation (INCT
GNSS-NavAer), funded by CNPq (National Council for Scientific and
Technological Development - process 465648/2014-2), FAPESP (Sao
Paulo Research Foundation - process 2017/01550-0) and CAPES
(Coordination for the Improvement of Higher Education Personnel).


%
%

\bibliographystyle{plain}
\bibliography{bib_scintillation}


\clearpage

\appendix
\section{Pseudorange measurements for position
recovery}\label{Appendix:PositioningError}
The pseudorange
observation model(at frequency ($f_i$) in the $L_i$ band can be
written as \citep{Langley1999,Marques2018},
\begin{equation}\label{eq:pseudorange}
P^{(s)} (\mathrm{r}) = \|\mathrm{r}-\mathrm{r}^{(s)}\|_2+\nu,
\end{equation}
where $P^s (\mathrm{r})$ is the measured pseudorange,
$\|\mathrm{r}-\mathrm{r}^{(s)}\|_2=\sqrt{(r_x-r_x^{(s)})^2+(r_y-r_y^{(s)})^2+(r_z-r_z^{(s)})^2}$
is the geometric range between the receiver's antenna at location
$\mathrm{r}=(r_x,r_y,r_z)$ and the satellite's antenna at location
$\mathrm{r}^{(s)}=(r^{(s)}_x,r^{(s)}_y,r^{(s)}_z)$ respectively.
Term $\nu$ represents the model errors and any unmodelled effects
including the clock errors. We note that in the current analysis, we
omit the clock uncertainty effect and thus to retrieve receiver's
position, $\mathrm{r}$, we need to have a set of at least three
equations of the form (eq.~\ref{eq:pseudorange}); in other words, at
least 3 satellites have to be available.

Since the observation model (eq.~\ref{eq:pseudorange}) is
non-linear, a first order Taylor approximation is used, i.e.
\begin{equation}\label{eq:taylor}
P^{(s)} (\mathrm{r}) = P^{(s)} (\mathrm{r}_0)+\left(\nabla
P^{(s)}(\mathrm{r}_0)\right)^\mathrm{T} \Delta\mathrm{r} +
\varepsilon^{(s)},
\end{equation}
where $P^{(s)} (\mathrm{r}_0)$ is an approximate (computed)
estimate, $\nabla P^\mathrm{(s)}(\mathrm{r}_0)= [\frac{\partial
P^{(s)}}{\partial r_x},\frac{\partial P^{(s)}}{\partial
r_y},\frac{\partial P^{(s)}}{\partial r_z}]^\mathrm{T}$,
$\Delta\mathrm{r} = [\Delta r_x, \Delta r_y, \Delta r_z]^\mathrm{T}$
is the position error and $\varepsilon^{(s)}$ includes all the
errors related to the numerical approximations and the model
uncertainties. Furthermore, based on (eq.~\ref{eq:pseudorange}),
$\nabla P^{(s)}(\mathrm{r}_0)
=[\frac{r_{x_0}-r_{x}^{(s)}}{\|\mathrm{r}_0-\mathrm{r}^{(s)}\|_2},\frac{r_{y_0}-r_{y}^{(s)}}{\|\mathrm{r}_0-\mathrm{r}^{(s)}\|_2},
\frac{r_{z_0}-r_{z}^{(s)}}{\|\mathrm{r}_0-\mathrm{r}^{(s)}\|_2}
]^\mathrm{T}$.

Then, we end up with a set of linear equations
\begin{equation}\label{eq:linearSystem_appendix}
b= A\Delta\mathrm{r}+\varepsilon,
\end{equation}
where  $b = [P^{(1)} (\mathrm{r}) -
P^{(1)}(\mathrm{r}_0),\ldots,P^{(S)} (\mathrm{r}) - P^{(S)}
(\mathrm{r}_0)]^\mathrm{T}\in\mathbb{R}^S$, $A=[\nabla
P^{(1)}(\mathrm{r}_0)\ldots\nabla
P^{(S)}(\mathrm{r}_0)]^\mathrm{T}\in\mathbb{R}^{S\times 3}$ and
$\varepsilon\in\mathbb{R}^S$. Here, we assume that even in the
absence of a loss of lock, scintillation can induce higher-order
errors in the observables that do not cancel out in dual frequency
combinations \citep{Aquino2009,IJssel2016}.%

\section{Defining the position dilution of
precision}\label{sec:insightInDilutionOfPrecision} { As we will show
in the following text, the position dilution of precision is a
measure of the uncertainty of the estimates $\widehat{\Delta
\mathrm{r}}$ and it is parameterized according to the statistical
characterization of the errors $\varepsilon$ and the linear model
used for the position estimates (eq.~\ref{eq:linearSystem}). In
particular,  by denoting $\Gamma_{xx}$, $\Gamma_{yy}$ and
$\Gamma_{zz}$ the diagonal elements of the general covariance
$\Gamma=(A^\mathrm{T}\Gamma_\varepsilon^{-1} A)^{-1}$
(eq.~\ref{eq:PostCovarianceWeights}), the standard deviations of
$\widehat{\Delta \mathrm{r}}$
(eq.~\ref{eq:generalExpectedPositioningError}) in X,Y and Z
direction are $\sigma_{x}=\sqrt{\Gamma_{xx}}$,
$\sigma_{y}=\sqrt{\Gamma_{yy}}$ and $\sigma_{z}=\sqrt{\Gamma_{zz}}$
respectively. The dilution of precision is defined as the normalized
(with a common scaling factor $\kappa$) root mean square (RMS) of
the standard deviations given by}
\begin{equation}
\mathrm{RMS}=\sqrt{\sigma_x^2+\sigma_y^2+\sigma_z^2}=\sqrt{\Gamma_{xx}+\Gamma_{yy}+\Gamma_{zz}}=\sqrt{\mathrm{tr}(\Gamma)}=
\sqrt{\mathrm{tr}((A^\mathrm{T}\Gamma_\varepsilon^{-1}
A)^{-1})}.\end{equation} So, the general form of the position
dilution of precision is
\begin{equation}\label{eq_generalPDOP}
\mathrm{PDOP}_{XYZ}=\frac{\mathrm{RMS}}{\sqrt{\kappa}} =
\frac{\sqrt{\mathrm{tr}((A^\mathrm{T}\Gamma_\varepsilon^{-1}
A)^{-1})}}{\sqrt{k}}= \sqrt{\mathrm{tr}((A^\mathrm{T}W A)^{-1})}
\end{equation}
where $W=\left(\frac{\Gamma_\varepsilon}{\kappa}\right)^{-1}$.
 Based on that ratio, we can see that the weights can be derived based on
 the prior knowledge about the error statistics.
Furthermore, from the expression of the $\mathrm{PDOP}_{XYZ}$  we
can understand that the dilution of precision is not the actual
positioning error, but a measure of the uncertainty of the estimates
$\widehat{\Delta \mathrm{r}}$. Therefore, the key feature of the
dilution of precision is that it is a single value that can be used
as an indicator about the reliability of a positioning system.
Moreover, we can notice that as matrix $A$ and $W$ do not depend on
the measurements, but only on the geometry and the weighting scheme,
dilution of precision can be computed from the satellite orbital
information  without needing real measurements. Currently, there are
empirical tables that relate PDOP values with the expected accuracy
of a positioning system (see for example table 1 in
\citep{Dutt2009}).

In the following text, we will see that the scaling factor $\kappa$
is defined based on the model of the observation error. More
precisely, for observation errors modeled as in
section~\ref{sec:standardPDOP}, the error covariance is given by
$\Gamma_\varepsilon = \gamma I^{S\times S}$ and
\begin{equation}\mathrm{RMS}=\sqrt{\gamma\mathrm{tr}(( A^\mathrm{T}
A)^{-1})}.\end{equation} Therefore, the common scaling factor is
$\kappa={\gamma}$ and $W=I^{S\times S}$. Then, we have the
well-known standard position dilution of precision given by
(eq.~\ref{eq:PDOP}).

For an observation error modelled as in
(eq.~\ref{eq:errordecomposition} of
section~\ref{sec:WPDOP_description}),
 the error covariance has two terms, i.e.
\begin{equation}\Gamma_\varepsilon=\gamma I^{S\times
S}+\Gamma_{\varepsilon_\mathrm{sc}}.\end{equation}
 Using the matrix
inversion lemma, we have that $\Gamma_\varepsilon^{-1}= \gamma^{-1}
(I^{S\times
S}-(\gamma\Gamma_{\varepsilon_\mathrm{sc}}^{-1}+I^{S\times
S})^{-1})$. Inserting the previous expression in the RMS, we have
that
\begin{equation}\mathrm{RMS}=\sqrt{\gamma\mathrm{tr}((A^\mathrm{T}U
A)^{-1})},\end{equation} where $U=I^{S\times
S}-(\gamma\Gamma_{\varepsilon_\mathrm{sc}}^{-1}+I^{S\times S})^{-1}$
and $\Gamma_\varepsilon^{-1}=\gamma^{-1} U$. Therefore, the scaling
factor $\kappa=\gamma$ and now
$W=
U$.  
Given the scintillation error covariance
$\Gamma_{\varepsilon_\mathrm{sc}}=\mathrm{diag}(h(r_\mathrm{ion}^{(s)}))$
we can easily show that
$W=\mathrm{diag}\left(\frac{\gamma}{\gamma+h(r_\mathrm{ion}^{(s)})}\right)
$ as in (eq.~\ref{eq:weights}).
Thefore, based on the previous error analysis, PDOP is given by
\begin{equation}\label{eq_generalPDOP_1}
\mathrm{PDOP}_{XYZ}=
\frac{\sqrt{\sigma_x^2+\sigma_y^2+\sigma_z^2}}{\sqrt{\gamma}}=\frac{\sqrt{\mathrm{tr}((A^{\mathrm{T}}\Gamma_\varepsilon^{-1}A)^{-1})}}{\sqrt{\gamma}}.
\end{equation}
%
%
%
\section{Average position error and position dilution of
precision}\label{sec:PDOPandPositionError} Let us first clarify that
$\Delta\mathrm{r}=\mathrm{r}_{\mathrm{tr}}-\mathrm{r}_0$ is the
position difference between the true XYZ position vector,
$\mathrm{r}_{\mathrm{tr}}=(r_{x_{\mathrm{tr}}},r_{y_{\mathrm{tr}}},r_{z_{\mathrm{tr}}})$,
and the computed (approximate) one,
$\mathrm{r}_0=(r_{x_0},r_{y_0},r_{z_0})$, and
$\widehat{\Delta\mathrm{r}}=\hat{\mathrm{r}}-\mathrm{r}_0$ is the
position vector difference between the estimated position vector
$\hat{\mathrm{r}}=(\hat{r}_{x},\hat{r}_{y},\hat{r}_{z})$ and
$\mathrm{r}_0$.

 Based on the linear system (eq.~\ref{eq:linearSystem}) and the
observation error modelled using a Gaussian distribution i.e.
$\varepsilon\sim\mathcal{N}(0,\Gamma_{\varepsilon})$, the estimated
position is given by
\begin{equation}\label{eq:positionEstimate_appendix}
\hat{\mathrm{r}}=\mathrm{r}_0+(A^{\mathrm{T}}\Gamma_\varepsilon^{-1}A)^{-1}A^{\mathrm{T}}\Gamma_\varepsilon^{-1}b.
\end{equation}

 The statistical expectation, denoted by $\mathbb{E}[\cdot] $ and
covariance, denoted by $\mathrm{Cov}[\cdot]$, of $\hat{\mathrm{r}}$
are
\begin{align}
\mathbb{E}[\hat{\mathrm{r}}]
=&\mathrm{r}_0+(A^{\mathrm{T}}\Gamma_\varepsilon^{-1}A)^{-1}A^{\mathrm{T}}\Gamma_\varepsilon^{-1}\mathbb{E}[b]=\mathrm{r}_{\mathrm{tr}}
\\
\mathrm{Cov}[\hat{\mathrm{r}}]=&
(A^{\mathrm{T}}\Gamma_\varepsilon^{-1}A)^{-1}A^{\mathrm{T}}\Gamma_\varepsilon^{-1}\mathrm{Cov}[b]\Gamma_\varepsilon^{-1}A
(A^{\mathrm{T}}\Gamma_\varepsilon^{-1}A)^{-1}=
(A^{\mathrm{T}}\Gamma_\varepsilon^{-1}A)^{-1}=\Gamma,
\end{align}
where
$\mathbb{E}[b]=\mathbb{E}[A(\mathrm{r}_{\mathrm{tr}}-\mathrm{r}_0)+\varepsilon]$,
$\mathbb{E}[\varepsilon]=0$ and
$\mathrm{Cov}[b]=\Gamma_\varepsilon$.

{ The position error vector between the true and the estimation
position vector is defined as
$\mathrm{e}=(e_x,e_y,e_z)=\hat{\mathrm{r}}-\mathrm{r}_{\mathrm{tr}}$.
}

{ The statistical expectation  and covariance of the position error
are respectively
\begin{align}
\mathbb{E}[\mathrm{e}]=&\mathbb{E}[\hat{\mathrm{r}}]-\mathrm{r}_{\mathrm{tr}}=0\\
\mathrm{Cov}[\mathrm{e}]=&\mathrm{Cov}[\hat{\mathrm{r}}]=\Gamma.
\end{align}
}

{ The average square magnitude of the position error, given by
$\|\mathrm{e}\|^2=e_x^2+e_y^2+e_z^2=(\hat{r}_x-r_{x_\mathrm{tr}})^2+(\hat{r}_y-r_{y_\mathrm{tr}})^2+(\hat{r}_z-r_{z_\mathrm{tr}})^2$,
is
\begin{equation}
\mathbb{E}[\|\mathrm{e}\|^2]=\mathbb{E}[e_x^2]+\mathbb{E}[e_y^2]+\mathbb{E}[e_z^2]=\mathrm{tr}(\mathrm{Cov}[\mathrm{e}])=
\mathrm{tr}(\Gamma)=\mathrm{tr}((A^{\mathrm{T}}\Gamma_\varepsilon^{-1}A)^{-1})
\end{equation}
From (eq.~\ref{eq_generalPDOP_1}) we get that
\begin{equation}
\mathbb{E}[\|\mathrm{e}\|^2]=\gamma\mathrm{PDOP}^2_{XYZ}
\end{equation}
where $\gamma$ is a scaling factor.} Finally,
\begin{equation}\label{eq:PDOP_prop_error}
\mathrm{PDOP}_{XYZ}\propto
\sqrt{\mathbb{E}[\|\mathrm{e}\|^2]}.\end{equation} Therefore, we can
see that the position dilution of precision is proportional to the
root of the expected square magnitude of the position error
\footnote{In reality, the true covariance of the observation errors,
denoted by $\Gamma_{\varepsilon}^{\mathrm{true}}$ is only partly
known. Of course, the better we can approximate the true covariance
$\Gamma_{\varepsilon}^{\mathrm{true}}$, the better we can predict
the expected error between the true and the estimated position by
using the PDOP. If the approximation $\Gamma_\varepsilon\approx
\Gamma_{\varepsilon}^{\mathrm{true}}$, we end up with the well-known
result (eq.~\ref{eq:PDOP_prop_error}). However, if
$\Gamma_\varepsilon \not \approx
\Gamma_{\varepsilon}^{\mathrm{true}}$, then
(eq.~\ref{eq:PDOP_prop_error}) is not valid anymore. For example,
during high scintillation activity we aim to improve the
approximation $\Gamma_\varepsilon\approx
\Gamma_{\varepsilon}^{\mathrm{true}}$ by introducing statistical
knowledge concerning scintillation into $\Gamma_\varepsilon$, and
thus
estimate a weighted PDOP that satisfies (eq.~\ref{eq:PDOP_prop_error}).}. 
We emphasize that the estimation of position error statistics would
require real measurements taken for a fixed matrix $A$ (i.e. for a
fixed satellite constellation) over a very long period of time. This
kind of requirement is in most cases far from feasible (obviously
because the available satellite constellations change over time)
which makes the model-based PDOP so crucial in quantifying the
expected position errors in positioning and navigation systems.
%
.

%
.

\end{document}